\documentclass[a4paper,10pt]{article}
\usepackage{amsmath}
\usepackage{amsfonts}
\usepackage{amssymb}
\usepackage{latexsym}
\usepackage{epsfig}
\usepackage{graphicx}
\usepackage{oldgerm}
\usepackage{theorem}

\def\Z{\mathbb{Z}}
\def\R{\mathbb{R}}
\def\C{\mathbb{C}}
\def\N{\mathbb{N}}
\def\H{\mathbb{H}}
\def\I{\mathbb{I}}

\def\bP{{\bf P}}
\def\bp{{\bf p}}

\def\cN{{\cal N}}
\def\cK{{\cal K}}
\def\cP{{\cal P}}
\def\cD{{\cal D}}
\def\cS{{\cal S}}

\def\RN{{\it R}_N}
\def\AN{{\it A}_{N-1}}
\def\BN{{\it B}_N}
\def\BNv{{\it B}^{\vee}_N}
\def\CN{{\it C}_N}
\def\CNv{{\it C}^{\vee}_N}
\def\BCN{{\it BC}_N}
\def\DN{{\it D}_N}

\def\z{\mib{z}}
\def\bZ{\mib{Z}}
\def\bxi{\boldsymbol{\xi}}

\def\mM{\mathfrak{M}}

\theorembodyfont{\itshape}

\newtheorem{thm}{Theorem}[section]
\newtheorem{lem}[thm]{Lemma}

\newtheorem{prop}[thm]{Proposition}

\newcommand{\mib}[1]{\mbox{\boldmath $#1$}}
\newcommand{\SSC}[1]{\section{#1}\setcounter{equation}{0}}
\newcommand{\qed}{\hbox{\rule[-2pt]{3pt}{6pt}}}

\begin{document}

\title{\bf 
Two-Dimensional Elliptic Determinantal \\
Point Processes and Related Systems
}
\author{
Makoto Katori
\footnote{
Fakult\"{a}t f\"{u}r Mathematik, Universit\"{a}t Wien, 
Oskar-Morgenstern-Platz 1, A-1090 Wien, Austria.
On sabbatical leave from
Department of Physics,
Faculty of Science and Engineering,
Chuo University, 
Kasuga, Bunkyo-ku, Tokyo 112-8551, Japan;
e-mail: katori@phys.chuo-u.ac.jp
}}
\date{20 November 2018}
\pagestyle{plain}
\maketitle

\begin{abstract}
We introduce new families of determinantal point processes (DPPs)
on a complex plane ${\mathbb{C}}$, which are classified into seven types
following the irreducible reduced affine root systems,
$R_N=A_{N-1}$, $B_N$, $B^{\vee}_N$, $C_N$, $C^{\vee}_N$, $BC_N$, $D_N$,
$N \in {\mathbb{N}}$. 
Their multivariate probability densities are doubly periodic
with periods $(L, iW)$, $0 < L, W < \infty$, $i=\sqrt{-1}$.
The construction is based on the orthogonality relations 
with respect to the double integrals over the fundamental domain,
$[0, L) \times i [0, W)$, which are proved in this paper 
for the $R_N$-theta functions introduced by Rosengren and Schlosser.
In the scaling limit $N \to \infty, L \to \infty$
with constant density
$\rho=N/(LW)$ and constant $W$, we obtain four types of
DPPs with an infinite number of
points on ${\mathbb{C}}$, which have periodicity with period $i W$. 
In the further limit $W \to \infty$ with constant $\rho$,
they are degenerated into three infinite-dimensional DPPs.
One of them is uniform on ${\mathbb{C}}$ and equivalent 
with the Ginibre point process
studied in random matrix theory, while
other two systems are rotationally symmetric around the origin,
but non-uniform on ${\mathbb{C}}$.
We show that the elliptic DPP
of type $A_{N-1}$ is identified with the particle section,
obtained by subtracting the background effect, 
of the two-dimensional exactly solvable model
for one-component plasma studied by Forrester.
Other two exactly solvable models of one-component plasma 
are constructed associated with the elliptic DPPs 
of types $C_N$ and $D_N$.
Relationship to the Gaussian free field on a torus
is discussed for these three exactly solvable plasma models.




\end{abstract}
\vspace{3mm}
\normalsize

\SSC
{Introduction} \label{sec:Introduction}

In a series of papers \cite{Kat15,Kat16,Kat17,Kat18},
we have studied elliptic extensions of 
determinantal point processes (DPPs) and determinantal processes (DPs),
which are realized in the systems of noncolliding Brownian motions (BM)
on a circle with radius $r >0$ or
in an interval $[0, \pi r]$ with appropriate boundary conditions
at the edges $x=0$ and $x=\pi r$.
These stochastic processes are defined in a finite time duration
$[0, t_{\ast}]$, $0 < t_{\ast} < \infty$,
in which the particle configurations at the final time
$t=t_{\ast}$ are pinned at specified configurations.
The basic idea of our elliptic extension is
based on the fundamental fact that
the Jacobi theta function $\vartheta_1(\xi; \tau)$ 
solves the following partial differential equation (PDE)
\begin{equation}
\frac{\partial^2 \vartheta_1(\xi; \tau)}{\partial \xi^2}
=4 \pi i \frac{\partial \vartheta_1(\xi; \tau)}{\partial \tau},
\label{eqn:PDE}
\end{equation}
where $i=\sqrt{-1}$. 
(Notations and formulas of the Jacobi theta functions used in this paper
are shown in Appendix \ref{sec:appendixA}.)
As functions of spatial and temporal coordinates
$x$ and $t$, we have parameterized the two variables
$\xi$ and $\tau$ as
\[
\xi=\xi(x)=\frac{x}{2 \pi r}, \quad
\tau=\tau(t)=\frac{it}{2 \pi r^2},
\]
$x \in \R, t \in [0, \infty)$. 
Then the PDE (\ref{eqn:PDE}) can be identified with the 
diffusion equation
\[
\left( \frac{\partial}{\partial t}- \frac{1}{2}
\frac{\partial^2}{\partial x^2} \right)
\vartheta_1(\xi(x); \tau(t))=0.
\]
With the positivity 
$\vartheta_1(\xi(x); \tau(t)) >0$
for $x \in (0, 2 \pi r), t > 0$,
this fact suggests that $\vartheta_1(\xi(x); \tau(t))$
will be used to describe probability laws of the systems of
BMs in the setting mentioned above.
Regarding the modular parameter $\tau$ as
an imaginary time and studying
time-evolution of a system by continuously changing $\tau$ may 
provide new applications of elliptic functions
and their related functions to stochastic analysis, 
but in our previous study,
only a situation with $\xi \in [0, 1]$ has been considered 
and quasi-double-periodicity of
$\vartheta_1(\xi; \tau)$ as a complex function of $\xi \in \C$
has not been used at all. 

In the present paper, we assume $0< L, W < \infty$ and consider
another parameterization, 
\begin{equation}
\xi= \xi(x, y)= \frac{z}{L} \quad
\mbox{with $z=x+i y$}, 
\qquad
\tau = i \frac{W}{L}, 
\label{eqn:xi_alpha1}
\end{equation}
for $(x, y) \in \R^2$.
It is obvious by this parameterization of $\xi$,
$\vartheta_1(\xi(x,y); \tau)$ is a harmonic function of
$(x, y) \in \R^2$.
The nontrivial point of this parameterization (\ref{eqn:xi_alpha1})
is found at $\tau$, which makes $\vartheta_1(\xi(x,y); \tau)$
have the following {\it quasi-double-periodicity} 
\begin{align*}
\vartheta_1(\xi(x+L, y); \tau)
&= - \vartheta_1(\xi(x,y); \tau), 
\nonumber\\
\vartheta_1(\xi(x, y+W); \tau)
&=-e^{-2\pi i \xi(x,y) - \tau \pi i} \vartheta_1(\xi(x,y); \tau).
\end{align*}
This fact suggests that, if we consider the absolute value
$|\vartheta_1(\xi(x,y); \tau)|$ with an appropriate
normalization, it will be used to describe
probability laws of suitable random point processes
defined on a complex plane $\C$ having a 
{\it fundamental domain} with periods $(L, i W)$, which is denoted as
\[
\Lambda_{(L, iW)}
\equiv 
[0, L) \times i [0, W) 
=\{z \in \C : 0 \leq \Re z < L, 0 \leq \Im z < W\}
\subset \C
\]
in this paper. The imaginary part of $\tau$,
$\Im \tau=W/L$, 
gives an {\it aspect ratio} of the rectangular shape
of this fundamental domain $\Lambda_{(L, i W)}$.

In order to obtain multivariate functions
which can be used to describe probability laws of DPPs and DPs,
the Macdonald denominator formulas
were used in the previous papers \cite{Kat15,Kat16,Kat17,Kat18}
and will be used in the present paper,
which were obtained by Rosengren and Schlosser \cite{RS06}
for the seven
types of irreducible reduced affine root systems,
$\RN=\AN$, $\BN, \BNv, \CN, \CNv, \BCN, \DN$,
$N \in \N$ \cite{Mac72}.
We introduce the following notations, 
\begin{align}
\Theta^{A}(\sigma, z, \tau) 
&=e^{2 \pi i \sigma z} 
\vartheta_2 (\sigma \tau + z; \tau),
\nonumber\\
\Theta^{B}(\sigma, z, \tau) 
&= e^{2 \pi i \sigma z} \vartheta_1 (\sigma \tau + z; \tau) 
- e^{-2 \pi i \sigma z} \vartheta_1 (\sigma \tau - z ; \tau), 
\nonumber\\
\Theta^{C}(\sigma, z, \tau) 
&= e^{2 \pi i \sigma z} \vartheta_2 (\sigma \tau + z; \tau)
- e^{-2 \pi i \sigma z} \vartheta_2 (\sigma \tau - z; \tau),
\nonumber\\
\Theta^{D}(\sigma, z, \tau)  
&= e^{2 \pi i \sigma z} \vartheta_2 (\sigma \tau + z; \tau)
+ e^{-2 \pi i \sigma z} \vartheta_2 (\sigma \tau - z; \tau),
\label{eqn:Theta}
\end{align}
for $\sigma \in \R, z \in \C$, 
$\tau \in \H \equiv \{z \in \C : \Im z > 0\}$.
With
\[
\sharp(\RN)
= \begin{cases}
A, \quad & \mbox{if $\RN=\AN$},
\cr
B, \quad & \mbox{if $\RN=\BN, \BNv$},
\cr
C, \quad & \mbox{if $\RN=\CN, \CNv, \BCN$},
\cr
D, \quad & \mbox{if $\RN=\DN$},
\end{cases}
\]
we set
\begin{equation}
M^{\RN}_j(z)
=\Theta^{\sharp(\RN)}(J(j)/\cN, \cN z/L, \cN \tau),
\label{eqn:setting1}
\end{equation}
for $\RN=\AN, \BN, \BNv, \CN, \CNv, \BCN, \DN$,
$j \in \{1,2, \dots, N\}$,
where $\cN=\cN^{\RN}$ and $J(j)=J^{\RN}(j)$
are defined by (\ref{eqn:N_R}) and (\ref{eqn:J_R}) given below, respectively.
In this setting, the Macdonald denominator formulas
of Rosengren and Schlosser (Proposition 6.1 in \cite{RS06})
are written as (\ref{eqn:Macdonald3}) with
(\ref{eqn:Macdonald_denominators}) in Section \ref{sec:Macdonald}. 

In the present paper, as a new aspect of the $\RN$-theta
functions of Rosengren and Schlosser, 
we prove the orthogonality relations for
the double integrals over the fundamental domain
$\Lambda_{(L, iW)}$,
\begin{equation}
\int_0^L dx \int_0^W dy \,
\exp \left( - \frac{2 \pi \cN}{LW} y^2 \right)
\overline{M^{\RN}_j(x+iy)}
M^{\RN}_k(x+iy)
= h^{\RN}_j \delta_{jk},
\label{eqn:d_ortho0}
\end{equation}
for $j, k \in \{1,2, \dots, N\}$, where $\overline{M^{\RN}_j(x+iy)}$
denotes the complex conjugate of $M^{\RN}_j(x+iy)$, 
and $\{h^{\RN}_j\}_{j=1}^N$ are given in 
Proposition \ref{thm:orthogonality_z}
depending on $\RN, N \in \N$.

Once such orthogonality relations are proved,
it is rather easy to construct DPPs on a complex plane $\C$.
We can also consider the scaling limit
$N \to \infty, L \to \infty$ with constant particle density 
$\rho=N/(LW)$ and constant $W$.
The obtained DPPs
with an infinite number of points are of four types and 
they are defined on $\C$
with a period $i W$ in the direction of imaginary axis.
In other words, these infinite dimensional DPPs essentially
live in a strip of height $W$ above the $x$-axis on $\C$. 
We study the further limit $W \to \infty$ with
constant $\rho$. In this limit,
we have three types of infinite DPPs on $\C$, one of which is 
identified with the {\it Ginibre point process}
studied in random matrix theory 
as the eigenvalue ensemble of
complex Gaussian random matrices
 \cite{Gin65,HKPV09,Shi15},
while other two infinite DPPs on $\C$ can be regarded as 
new examples of the {\it Mittag--Leffler fields} \cite{AK13,AKM18,AKS18}. 

Appearance of the exponential weights
$\exp(-2 \pi \cN y^2/(LW))$ in 
the double integrals (\ref{eqn:d_ortho0})
is essential, and we derive them by 
requiring that the probability measures of
the DPPs on $\C$ should be {\it doubly periodic} 
with respect to all $N$ complex variables
$(z_1, \dots, z_N)=
(x_1+i y_1, \dots, x_N+i y_N)$ 
which describe $N$-point configurations on $\C$.

In \cite{For06}, Forrester studied a particle system
in $\Lambda_{(L, i W)}$ with doubly periodic boundary conditions
such that $N$ mobile particles, each of which
is charged +1, are interacting via the pair potential 
given using logarithmic of $\vartheta_1(z/L; \tau)$
and these particles are confined to the domain $\Lambda_{(L, iW)}$.
He assumed that a uniform background with negative charge
density $-N/LW$ exists and the system is
neutralized. Such a system
consisting of positively charged $N$ particles
and negatively charged background is called
a {\it one-component plasma model} \cite{JT96,For06,For10}.
In the present paper, we show that the elliptic DPP
of type $\AN$ is identified with 
the particle section of 
Forrester's one-component plasma model,
in which the background effect is subtracted.
We also show that other two elliptic DPPs
of types $\CN$ and $\DN$ are 
also realized as the particle sections of one-component
plasma models, while these additional two systems are not
perfectly neutralized.
This consideration gives another derivation
of the exponential weights
$\exp(-2 \pi \cN y^2/(LW))$ in (\ref{eqn:d_ortho0})
at least for the three systems
of types $\AN$, $\CN$, and $\DN$.
Forrester claimed the equivalence between 
{\it exact-solvability} and {\it double-periodicity}
in his plasma model of type $\AN$.
This statement is extended to all seven types though 
the Macdonald denominator formulas of 
Rosengren and Schlosser \cite{RS06} and the 
orthogonality relations (\ref{eqn:d_ortho0}).

Forrester discussed an interesting relationship
of his one-component plasma model
to the {\it Gaussian free field} (GFF) defined on a torus
studied by Cardy \cite{Car90}.
We develop his argument to our two
additional models of types $\CN$ and $\DN$. 
Correspondence to 
the modular invariance of
the partition function 
discussed by Cardy for the GFF on a torus,
we find the correction terms in
large $N$ expansion of the free energies
of our new plasma models of types $\CN$ and $\DN$,
which are invariant under the transformation of
the aspect ratio $W/L \to L/W$.

The paper is organized as follows.
In Section \ref{sec:preliminaries}, we first introduce notations
used in this paper 
associated with Appendix \ref{sec:appendixA}, 
and we list out the Macdonald denominator formulas 
of Rosengren and Schlosser \cite{RS06}
in our setting (\ref{eqn:setting1}). 
Then we construct the seven types of
probability weights, $Q^{\RN}(\z)$ for
$\RN=\AN$, $\BN$, $\BNv$, $\CN$, $\CNv$, $\BCN$, $\DN$,
$N \in \N$, 
which are all doubly periodic with periods $(L, i W)$
with respect to $N$-component complex variables $\z$ 
representing point configurations for the systems. 
The orthogonality relations (\ref{eqn:d_ortho0}) 
are proved in Proposition \ref{thm:orthogonality_z}
for $\{M^{\RN}_j(z)\}_{j=1}^N$ given by (\ref{eqn:setting1}). 
In Section \ref{sec:DPP}, the seven types of
point processes on $\C$,
$(\Xi^{\RN}, \bP^{\RN})$,
$\RN=\AN$, $\BN$, $\BNv$, $\CN$, $\CNv$, $\BCN$, $\DN$,
are constructed by properly normalizing the 
doubly periodic weights
$Q^{\RN}(\z)$ to define the probability laws
$\bP^{\RN}$. Then we can prove that
all of them are {\it determinantal}
following the standard method in 
{\it random matrix theory}
\cite{Meh04,For10,AGZ10,Kat15_Springer},
and we give the {\it correlation kernels}
using the orthogonal theta functions,
$\{M^{\RN}_j(z)\}_{j=1}^N$ 
(Theorem \ref{thm:mainA1}).
The scaling limits 
$N \to \infty, L \to \infty$
with constant point-density $\rho$ and $W$ are calculated
for the correlation kernels 
(Proposition \ref{thm:infinite_systems1}), 
and the four types of DPPs
with an infinite number of particles are derived on $\C$ 
(Theorem \ref{thm:scaling_limits}).
In the further limit $W\to \infty$ with
$\rho=$ const., 
these four systems are degenerated into
three infinite DPPs (Theorem \ref{thm:Ginibre}). 
In Section \ref{sec:plasma_GFF}, we first review the one-component
plasma model studied by Forrester \cite{For06}
and then introduce other two models,
which are defined in $\Lambda_{(L, iW)}$ with
doubly periodic conditions.
We show that the particle section of
Forrester's model is identified with 
the elliptic DPP of type $\AN$, $(\Xi^{\AN}, \bP^{\AN})$.
We also prove that other two models realize 
$(\Xi^{\CN}, \bP^{\CN})$ and $(\Xi^{\DN}, \bP^{\DN})$
as their particle sections
(Theorem \ref{thm:solvable_plasma}).
Relationship to the GFF defined on a torus
studied by Cardy \cite{Car90} is discussed
for these three types of exactly solvable plasma models.
Appendices \ref{sec:appendixB} and \ref{sec:appendixC} 
support this section. 
Section \ref{sec:remarks} is devoted to concluding remarks.

\SSC
{Preliminaries} \label{sec:preliminaries}
\subsection{Macdonald denominator formulas
of Rosengren and Schlosser}
\label{sec:Macdonald}

Assume that $N \in \N \equiv \{1,2, \dots\}$.
As extensions of the Weyl denominators for
classical root systems, Rosengren and Schlosser \cite{RS06}
studied the Macdonald denominators \cite{Mac72} for the seven
types of irreducible reduced affine root systems,
$W^{\RN}(\bxi)$, $\bxi=(\xi_1, \dots, \xi_N) \in \C^N$,
$\RN=\AN$, $\BN, \BNv, \CN, \CNv, \BCN, \DN$,
$N \in \N$. See also \cite{Kra05,War02}.
Up to trivial factors they are written using the Jacobi theta functions
as follows.
\begin{align}
W^{\AN}(\bxi; \tau) &=
\prod_{1 \leq j < k \leq N} \vartheta_1(\xi_k-\xi_j; \tau),
\nonumber\\
W^{\BN}(\bxi; \tau) &=
\prod_{\ell=1}^N \vartheta_1(\xi_{\ell}; \tau)
\prod_{1 \leq j < k \leq N} \Big\{
\vartheta_1(\xi_k-\xi_j; \tau) \vartheta_1(\xi_k+\xi_j; \tau) \Big\},
\nonumber\\
W^{\BNv}(\bxi; \tau) &=
\prod_{\ell=1}^N \vartheta_1(2 \xi_{\ell}; 2 \tau)
\prod_{1 \leq j < k \leq N} \Big\{
\vartheta_1(\xi_k-\xi_j; \tau) \vartheta_1(\xi_k+\xi_j; \tau) \Big\},
\nonumber\\
W^{\CN}(\bxi; \tau) &=
\prod_{\ell=1}^N \vartheta_1(2 \xi_{\ell}; \tau)
\prod_{1 \leq j < k \leq N} \Big\{
\vartheta_1(\xi_k-\xi_j; \tau) \vartheta_1(\xi_k+\xi_j; \tau) \Big\},
\nonumber\\
W^{\CNv}(\bxi; \tau) &=
\prod_{\ell=1}^N \vartheta_1 \left(\xi_{\ell}; \frac{\tau}{2} \right)
\prod_{1 \leq j < k \leq N} \Big\{
\vartheta_1(\xi_k-\xi_j; \tau) \vartheta_1(\xi_k+\xi_j; \tau) \Big\},
\nonumber\\
W^{\BCN}(\bxi; \tau) &=
\prod_{\ell=1}^N \Big\{ \vartheta_1(\xi_{\ell}; \tau) \vartheta_0(2 \xi_{\ell}; 2 \tau) 
\Big\}
\prod_{1 \leq j < k \leq N} \Big\{
\vartheta_1(\xi_k-\xi_j; \tau) \vartheta_1(\xi_k+\xi_j; \tau) \Big\},
\nonumber\\
W^{\DN}(\bxi; \tau) &=
\prod_{1 \leq j < k \leq N} \Big\{
\vartheta_1(\xi_k-\xi_j; \tau) \vartheta_1(\xi_k+\xi_j; \tau) \Big\},
\label{eqn:Macdonald_denominators}
\end{align}
where $\tau \in \H$.
In the present paper, we use 
the $\AN$-theta functions of
norm $t=e^{2 \pi i \widetilde{t}_N}$ with
\[
\widetilde{t}_N= \begin{cases}
N \tau/2, & \mbox{if $N$ is even},
\cr
(1+ N \tau)/2, & \mbox{if $N$ is odd},
\end{cases}
\]
and the $\RN$-theta functions, 
for $\RN=\BN, \BNv, \CN, \CNv, \BCN, \DN$,
of Rosengren and Schlosser \cite{RS06}, 
which are parameterized as (\ref{eqn:setting1}).
Here 
\begin{equation}
\cN=\cN^{\RN} = \begin{cases}
N, \quad & \RN = \AN, \\
2N-1, \quad & \RN = \BN, \\
2N, \quad & \RN = \BNv, \CNv, \\
2(N+1), \quad & \RN =\CN, \\
2N+1, \quad & \RN = \BCN, \\
2(N-1), \quad & \RN = \DN,
\end{cases}
\label{eqn:N_R}
\end{equation}
and
\begin{equation}
J(j)=J^{\RN}(j) = \begin{cases}
j-1/2, \quad & \RN = \AN, \CNv,
\\
j-1, \quad & \RN=\BN, \BNv, \DN,
\\
j, \quad & \RN=\CN, \BCN.
\end{cases}
\label{eqn:J_R}
\end{equation}
The explicit expressions of these functions are
given as following, 
\begin{align}
& M^{\AN}_j(z) = M^{\AN}_j(z; L, W)
\nonumber\\
& \quad 
=e^{2 \pi i J(j) z/L} 
\vartheta_2 \Big(J(j) \tau + \cN z/L; \cN \tau \Big),
\nonumber\\
& M^{\RN}_j(z) = M^{\RN}_j(z; L, W)
\nonumber\\
& \quad 
= e^{2 \pi i J(j) z/L} \vartheta_1 \Big(J(j) \tau + \cN z/L; \cN \tau \Big)
- e^{-2 \pi i J(j) z/L} \vartheta_1 \Big(J(j) \tau - \cN z/L; \cN \tau \Big), 
\nonumber\\
& \hskip 10cm 
\mbox{for $\RN=\BN, \BNv$},
\nonumber\\
& M^{\RN}_j(z) = M^{\RN}_j(z; L, W)
\nonumber\\
& \quad
= e^{2 \pi i J(j) z/L} \vartheta_2 \Big(J(j) \tau + \cN z/L; \cN \tau \Big) 
- e^{-2 \pi i J(j) z/L} \vartheta_2 \Big(J(j) \tau - \cN z/L; \cN \tau \Big),
\nonumber\\
& \hskip 10cm
\mbox{for $\RN=\CN, \CNv, \BCN$},
\nonumber\\
&M^{\DN}_j(z) = M^{\DN}_j(z; L, W)
\nonumber\\
& \quad 
= e^{2 \pi i J(j) z/L} \vartheta_2 \Big(J(j) \tau + \cN z/L; \cN \tau \Big)
+ e^{-2 \pi i J(j) z/L} \vartheta_2 \Big(J(j) \tau - \cN z/L; \cN \tau \Big),
\label{eqn:MA_D1}
\end{align}
where $\tau=i W/L$.

Let $\eta(\tau)$ be the {\it Dedekind modular function} 
(see, for instance, Sec.23.15 in \cite{NIST10}),
\begin{equation}
\eta(\tau)= e^{\tau \pi i/12} \prod_{n=1}^{\infty} (1-e^{2n \tau \pi i}).
\label{eqn:Dedekind1}
\end{equation}
In the present setting (\ref{eqn:setting1}), 
the Macdonald denominator formulas of
Rosengren and Schlosser (Proposition 6.1 in \cite{RS06})
are written as follows.
\begin{align}
& \det_{1 \leq j, k \leq N}
\Big[ M^{\AN}_j(z_k) \Big]
\nonumber\\
& \quad
=\begin{cases}
\displaystyle{
i^{N/2} a(\tau)
\vartheta_0 \left(\sum_{j=1}^N z_{j}/L ; \tau \right)
W^{\AN}(\z/L; \tau)
},
& \mbox{if $N$ is even},
\cr
\displaystyle{
i^{-(N-1)/2} a(\tau)
\vartheta_3 \left(\sum_{j=1}^N x_{j}/L ; \tau \right)
W^{\AN}(\z/L; \tau)
},
& \mbox{if $N$ is odd},
\end{cases}
\nonumber\\
& \det_{1 \leq j, k \leq N}
\Big[ M^{\RN}_j(z_k) \Big]
= a(\tau)
W^{\RN}(\z/L; \tau),
\quad \mbox{for $\RN=\BN, \BNv, \DN$},
\nonumber\\
& \det_{1 \leq j, k \leq N}
\Big[ M^{\RN}_j(z_k) \Big]
= i^{-N} a(\tau)
W^{\RN}(\z/L; \tau),
\quad \mbox{for $\RN=\CN, \CNv, \BCN$},
\label{eqn:Macdonald3}
\end{align}
where $a(\tau) \in \R$ are given by
\begin{equation}
a(\tau)=a^{\RN}(\tau)=
\begin{cases}
e^{-(2N-1)(2N+1) \tau \pi i/12} 
\eta(\tau)^{-(N-1)(N-2)/2},
& \RN=\AN, 
\nonumber\\
2 e^{-N(N-1) \tau \pi i/6} 
\eta(\tau)^{-N(N-1)},
& \RN=\BN, 
\nonumber\\
2 e^{-(N-1)(2N-1) \tau \pi i/12} 
\eta(\tau)^{-(N-1)^2} \eta(2 \tau)^{-(N-1)},
& \RN=\BNv, 
\nonumber\\
e^{-N(2N+1) \tau \pi i/12} 
\eta(\tau)^{-N(N-1)},
& \RN=\CN, 
\nonumber\\
e^{-(2N-1)(2N+1) \tau \pi i/24} 
\eta(\tau)^{-(N-1)^2} \eta(\tau/2)^{-(N-1)},
& \RN=\CNv,
\nonumber\\
e^{-N(N+1) \tau \pi i/6} 
\eta(\tau)^{-N(N-1)} \eta(2 \tau)^{-N},
& \RN=\BCN, 
\nonumber\\
4 e^{- N(2N-1) \tau \pi i/12} 
\eta(\tau)^{-N(N-2)},
& \RN=\DN,
\end{cases}
\label{eqn:a1}
\end{equation}
and $\z/L=(z_1/L, \dots, z_N/L)$ with $\z =(z_1, \dots, z_N) \in \C^N$.

\subsection{Doubly periodic weights}
\label{sec:weights}

For $z \in \C$, we write $x=\Re z$ and $y=\Im z$.
Let $s(N)=0$ if $N$ is even, and $s(N)=3$ if $N$ is odd.
For $\z=(z_1, \dots, z_N) \in \C^N$,
define
\begin{align}
&C(\z)=C^{\RN}(\z; L, W)
\nonumber\\
& \qquad
=\begin{cases}
\displaystyle{\exp \left( 
- \frac{\pi \cN}{LW} \sum_{j=1}^N y_j^2 \right)
\vartheta_{s(N)} \left(
\sum_{k=1}^N \frac{z_k}{L}; \tau \right),
}
& \mbox{for $\RN=\AN$},
\cr
\displaystyle{\exp \left( 
- \frac{\pi \cN}{LW} \sum_{j=1}^N y_j^2 \right),
}
& \mbox{for $\RN=\BN, \BNv, \CN, \CNv, \BCN, \DN$}.
\end{cases}
\label{eqn:CRN1}
\end{align}
For $\z=(z_1, \dots, z_N) \in \C^N$, we consider the
shift operators $\sigma_m(w)$ with
$m=1,2, \dots, N$, $w \in \C$ such that, 
for a function $f$ of $\z$,
\[
\sigma_m(w) f(\z) =
f(z_1, \dots, z_{m-1}, z_m+w, z_{m+1}, \dots, z_N).
\]
Then it is easy to verify the following. 
\begin{lem}
\label{thm:qRN1}
For $\RN=\AN, \BN, \BNv, \CN, \CNv, \BCN, \DN$, define
\[
q(\z)=q^{\RN}(\z)=
C(\z)
W^{\RN} (\z/L; \tau).
\]
Then these multivariate functions 
$q(\z)$ are quasi-double-periodic 
in the sense that, for $m=1,2, \dots, N$, 
\begin{align*}
q(\sigma_m(L) \z) &= {\rm sgn}_{(L)} q(\z),
\nonumber\\
q(\sigma_m(iW) \z) &= {\rm sgn}_{(i W)}
e^{- 2 \pi i \cN x_m/L} q(\z),
\end{align*}
where
\begin{align*}
{\rm sgn}_{(L)} &={\rm sgn}_{(L)}^{\RN} = \begin{cases}
1, & \mbox{for $\RN=\AN$ with $N$ odd, $\BNv, \CN, \DN$},
\cr
-1, & \mbox{for $\RN=\AN$ with $N$ even, $\BN, \CNv, \BCN$},
\end{cases}
\nonumber\\
{\rm sgn}_{(i W)} &=
{\rm sgn}_{(i W)}^{\RN} = \begin{cases}
1, & \mbox{for $\RN=\AN, \CN, \CNv, \BCN, \DN$},
\cr
-1, & \mbox{for $\RN=\BN, \BNv$}.
\end{cases}
\end{align*}
\end{lem}
\vskip 0.3cm
We write the complex conjugate of $z=x+i y \in \C$ as
$\overline{z}=x-i y$ and 
$\overline{\z} \equiv (\overline{z_1}, \dots, \overline{z_N})$. 
By the definitions (\ref{eqn:Macdonald_denominators})
and (\ref{eqn:CRN1}), and by parity (\ref{eqn:even_odd})
of the Jacobi theta functions, 
\begin{align*}
\overline{W^{\RN}\left(\z/L; \tau \right)}
&= W^{\RN} \left( \overline{\z}/L); \tau \right),
\nonumber\\
\overline{ C^{\AN}(\z) }
&= \exp \left( - \frac{\pi \cN}{LW}
\sum_{j=1}^N y_j^2 \right)
\vartheta_{s(N)} \left( \sum_{k=1}^N \frac{\overline{z_k}}{L}; \tau \right),
\nonumber\\
\overline{ C^{\RN}(\z) }
&= C^{\RN}(\z) 
=\exp \left( - \frac{\pi \cN}{LW}
\sum_{j=1}^N y_j^2 \right),
\quad \RN=\BN, \BNv, \CN, \CNv, \BCN, \DN.
\end{align*}

Now we define the real-valued functions as follows,
\begin{align}
&Q(\z) = Q^{\RN}(\z)
\equiv |q(\z) |^2
=\overline{q(\z)} q(\z)
\nonumber\\
&=\begin{cases}
\displaystyle{
\exp \left( - \frac{2 \pi \cN}{LW}
\sum_{j=1}^N y_j^2 \right)
\left| \vartheta_{s(N)} \left(
\sum_{k=1}^N \frac{z_k}{L} ; \tau \right) \right|^2
\left| W^{\AN} \left( \z/L; \tau \right) \right|^2
}, 
& \mbox{$\RN=\AN$},
\cr
\displaystyle{
\exp \left( - \frac{2 \pi\cN}{LW}
\sum_{j=1}^N y_j^2 \right)
\left| W^{\RN} \left(\z/L; \tau \right) \right|^2
}, &
\mbox{$\RN=\BN, \BNv, \CN, \CNv, \BCN, \DN$}.
\end{cases}
\label{eqn:QRN1}
\end{align}
Then the following is immediately concluded from Lemma \ref{thm:qRN1}.

\begin{prop}
\label{thm:totally_elliptic}
The multivariate functions $Q(\z)$ defined by
{\rm (\ref{eqn:QRN1})} are doubly periodic 
with periods $(L, iW)$ in the sense that
they satisfy the equalities
\[
\sigma_m(L) Q(\z) = Q(\z),
\qquad
\sigma_m(iW) Q(\z) = Q(\z),
\]
for any $m=1,2, \dots, N$, 
\end{prop}

\subsection{Orthogonality}
\label{sec:orthogonality}

In Lemma 2.1 in \cite{Kat18}, we proved the orthogonality 
for a version,
\[\Theta^{\sharp(\RN)}(J(j)/\cN, \cN x/(2\pi r), i \cN^2 t/(2 \pi r^2)),
\quad 
(x, t) \in \R  \times [0, \infty), 
\quad j \in \{1,2, \dots, N \}, 
\]
of the seven series of $\RN$-theta functions of
Rosengren and Schlosser \cite{RS06}.
By the similar argument, we can prove the following orthogonality
for the present version
$\{M^{\RN}_j(z)\}_{j=1}^N$ defined by (\ref{eqn:setting1}) for
a complex variable
$z=x+iy, (x, y) \in \R^2$, 
respect to the integral over $x \in [0, L)$.
Here we note that the following relations hold for the
complex conjugates of $\{M^{\RN}_j(z)\}_{j=1}^N, z \in \C$;
\begin{align*}
\overline{M^{\AN}_j(z)} &= M^{\AN}_j(- \overline{z}),
\nonumber\\
\overline{M^{\RN}_j(z)} &= -M^{\RN}_j(-\overline{z}) 
= M^{\RN}(\overline{z}),
\quad \mbox{for $\RN=\BN, \BNv$},
\nonumber\\
\overline{M^{\RN}_j(z)} &= M^{\RN}_j(-\overline{z}) 
= - M^{\RN}(\overline{z}),
\quad \mbox{for $\RN=\CN, \CNv, \BCN$},
\nonumber\\
\overline{M^{\DN}_j(z)} &= M^{\DN}_j(-\overline{z}) 
= M^{\DN}(\overline{z}).
\end{align*}

\begin{lem}
\label{thm:orthogonality_x}
Let $z=x+i y$, $(x, y) \in \R^2$.
For $\RN=\AN$,$\BN$,$\BNv$,$\CN$,$\CNv$,$\BCN$,$\DN$, 
if $j, k \in \{1,2, \dots, N\}$, 
\[
\int_0^{L} \overline{M^{\RN}_j(z)}
M^{\RN}_k(z) dx
= m^{\RN}_j(y) \delta_{jk},
\]
where
\begin{align}
m^{\AN}_j(y) &=
L e^{-4 \pi J(j) y/L} 
\vartheta_2 \Big( 2 (J(j) \tau +  i \cN y/L ) ; 
2 \cN \tau \Big),
\quad \quad j \in \{1,2, \dots, N\}, 
\nonumber\\
m^{\RN}_j(y) &=
L \left\{ e^{4 \pi J (j) y/L} 
\vartheta_2 \Big( 2 (J(j) \tau -  i \cN y/L) ; 
2 \cN \tau \Big) \right.
\nonumber\\
& \quad \quad
\left. + 
e^{-4 \pi J(j) y/L} 
\vartheta_2 \Big( 2 (J (j) \tau +  i \cN y/L)  ; 
2 \cN \tau \Big) \right\}, 
\quad j \in \{1,2, \dots, N\},
\nonumber\\
& \hskip 8cm
\quad \mbox{for $\RN=\CN, \CNv, \BCN$},
\nonumber\\
m^{\RN}_j(y) &= \begin{cases}
4 L \vartheta_2 \Big( 2 i \cN y/L) ; 
2 \cN \tau \Big),
& j=1, 
\\
L \left\{ e^{4 \pi J (j) y/L} 
\vartheta_2 \Big( 2 (J (j) \tau - i \cN y/L) ; 
2 \cN \tau \Big) \right.
&
\\
\quad \left.
+ e^{-4 \pi J(j) y/L} 
\vartheta_2 \Big( 2 (J(j) \tau + i \cN y/L) ; 
2 \cN \tau \Big) \right\},
& j \in \{2, 3, \dots, N\},
\end{cases}
\nonumber\\
& \hskip 9cm \mbox{for $\RN=\BN, \BNv$},
\nonumber\\
m^{\DN}_j(y) &= \begin{cases}
4 L 
\vartheta_2 \Big( 2 i \cN y/L; 2 \cN \tau \Big),
& j=1,
\\
L \left\{ e^{4 \pi J(j) y/L} 
\vartheta_2 \Big( 2 (J(j) \tau -i \cN y/L) ; 
2 \cN \tau \Big) \right.
&
\\
\quad \left.
+ e^{-4 \pi J(j) y/L} 
\vartheta_2 \Big( 2 (J(j) \tau + i \cN y/L) ; 
2 \cN \tau \Big) \right\},
& j \in \{2, 3, \dots, N-1 \},
\\
2L \left\{ e^{4 \pi J(N) y/L} 
\vartheta_2 \Big( 2 (J(N) \tau - i \cN y/L); 
2 \cN \tau \Big) \right.
&
\\
\quad \left.
+ e^{-4 \pi J(N) y/L} 
\vartheta_2 \Big( 2 (J(N) \tau + i \cN y/L); 
2 \cN \tau \Big) \right\},
& j =N.
\end{cases}
\label{eqn:ortho_const_y}
\end{align}
\end{lem}

In the present paper, we prove the following orthogonality
for the version (\ref{eqn:setting1}) 
of the $\RN$-theta functions of Rosengren and Schlosser \cite{RS06}
with respect to the double integrals
over the fundamental domain $\Lambda_{(L, iW)}$. 

\begin{prop}
\label{thm:orthogonality_z}
Let $z=x+i y, (x, y) \in \R^2$.
If $j, k \in \{1,2, \dots, N\}$, then
for $\RN=\AN$, $\BN$, $\BNv$, $\CN$, $\CNv$, $\BCN$, $\DN$, 
\begin{equation}
\int_0^{L} dx \int_0^W dy \,
\exp \left( - \frac{2 \pi \cN}{LW} y^2 \right)
\, \overline{M^{\RN}_j(z)}
M^{\RN}_k(z) 
= h^{\RN}_j\delta_{jk},
\label{eqn:ortho_z_RN}
\end{equation}
where
\begin{align}
h^{\AN}_j 
&= 
\frac{LW}{\sqrt{2 \cN \Im \tau}}
e^{- 2 \tau \pi i J(j)^{2}/\cN},
\quad j \in \{1,2, \dots, N\},
\nonumber\\
h^{\RN}_j 
&= 
\frac{2LW}{\sqrt{2 \cN \Im \tau}}
e^{- 2 \tau \pi i J(j)^{2}/\cN},
\quad j \in \{1,2, \dots, N\},
\quad \mbox{for $\RN=\CN, \CNv, \BCN$},
\nonumber\\
h^{\RN}_j 
&= \begin{cases}
\displaystyle{\frac{4LW}{\sqrt{2 \cN \Im \tau}}}
e^{- 2 \tau \pi i J(j)^{2}/\cN},
\quad j=1,
\cr
& \cr
\displaystyle{\frac{2LW}{\sqrt{2 \cN \Im \tau}}}
e^{- 2 \tau \pi i J(j)^{2}/\cN},
\quad j \in \{2,3, \dots, N\}, 
\end{cases}
\quad \mbox{for $\RN=\BN, \BNv$},
\nonumber\\
h^{\DN}_j 
&= \begin{cases}
\displaystyle{\frac{4LW}{\sqrt{2 \cN \Im \tau}}}
e^{- 2 \tau \pi i J (j)^{2}/\cN},
\quad j \in \{1, N \}, 
\cr
& \cr
\displaystyle{\frac{2LW}{\sqrt{2 \cN \Im \tau}}}
e^{- 2 \tau \pi i J(j)^{2}/\cN},
\quad j \in \{2,3, \dots, N-1 \}.
\end{cases}
\label{eqn:ortho_const_z}
\end{align}
\end{prop}
\vskip 0.3cm
\noindent{\it Proof} \,
By Lemma \ref{thm:orthogonality_x},
the orthogonality (\ref{eqn:ortho_z_RN}) holds with
\[
h^{\RN}_j=\int_0^W 
e^{-2 \pi i \cN (y/L)^2/\tau} m^{\RN}_j(y) dy,
\quad j=1,2, \dots, N.
\]
If we apply Jacobi's imaginary transformation 
(\ref{eqn:Jacobi_imaginary}), we obtain the equality
\begin{align*}
& \vartheta_2 \left( 2
 \left(J(j) \tau \pm i \frac{\cN y}{L} \right); 2 \cN \tau \right)
\nonumber\\
& \quad
=\frac{1}{\sqrt{2 \cN \Im \tau}}
e^{- 2 \tau \pi i J(j)^2/\cN \pm 4 \pi J(j) y/L
+ 2 \pi i \cN (y/L)^2/\tau}
\vartheta_0 \left(
\frac{J(j)}{\cN} \pm i \frac{y}{L\tau};
- \frac{1}{2 \cN \tau} \right).
\end{align*}
Hence
\begin{align*}
& e^{-2 \pi i \cN (y/L)^2/\tau}
e^{\mp 4 \pi J(j) y/L}
\vartheta_2 \left( 2
\left( J(j) \tau \pm i \frac{\cN y}{L} \right); 2 \cN \tau \right)
\nonumber\\
& \quad
= \frac{1}{\sqrt{2 \cN \Im \tau}}
e^{- 2 \tau \pi i J(j)^2/\cN}
\vartheta_0 \left(
\frac{J(j)}{\cN} \pm i \frac{y}{L \tau};
- \frac{1}{2 \cN \tau} \right).
\end{align*}
By the definition of $\vartheta_0$ given by (\ref{eqn:theta}),
\[
\int_0^W 
\vartheta_0 \left(
\frac{J(j)}{\cN} \pm i \frac{y}{L \tau};
-\frac{1}{2 \cN \tau} \right) dy
= \sum_{n \in \Z} (-1)^n
e^{-n^2 \pi i/(2 \cN \tau)
+ 2 n \pi J(j)/\cN}
\int_0^W e^{\pm 2 n \pi i y/W} dy.
\]
Since $\int_0^W e^{\pm 2 n \pi i y/W} dy=W \delta_{n 0}$,
the above integral is equal to $W$.
Then, from (\ref{eqn:ortho_const_y}) in
Lemma \ref{thm:orthogonality_x}, 
(\ref{eqn:ortho_const_z}) are derived.
The proof is complete. \qed
\vskip 0.3cm

\SSC
{Elliptic Determinantal Point Processes
on a Complex Plane} \label{sec:DPP}
\subsection{DPPs with a finite number of points}
\label{sec:finite}

Combination of the Macdonald denominator formulas
(\ref{eqn:Macdonald3}) with (\ref{eqn:a1})
and the orthogonality (Proposition \ref{thm:orthogonality_z})
for the present version (\ref{eqn:setting1})
of the $\RN$-theta functions
of Rosengren and Schlosser \cite{RS06}, 
the following multiple integral formulas
are derived for the doubly periodic weight functions
$Q^{\RN}(\z)$ defined by (\ref{eqn:QRN1}).

\begin{lem}
\label{thm:Qint}
For $\RN=\AN$, $\BN$, $\BNv$, $\CN$, $\CNv$, $\BCN$, $\DN$, 
\begin{equation}
\frac{1}{N!}
\int_{{\Lambda_{(L, iW)}}^N} 
Q^{\RN}(\z) d \z
\equiv 
\frac{1}{N!}
\prod_{j=1}^N \int_0^L d x_j \int_0^W d y_j \,
Q^{\RN}(\z)
=Z^{\RN},
\label{eqn:Qint1}
\end{equation}
where
\begin{equation}
Z^{\RN}=Z^{\RN}(L, W)
=2^{\delta} \frac{(LW)^N}{(\cN \Im \tau)^{N/2}}
\eta(\tau)^{\kappa} g(\tau),
\label{eqn:Z1}
\end{equation}
with
\begin{align}
\delta &= \delta^{\RN}=
\begin{cases}
-N/2, & \RN=\AN, \cr
(N-2)/2, & \RN=\BN, \BNv, \cr
N/2, & \RN=\CN, \CNv, \BCN, \cr
(N-4)/2, & \RN=\DN,
\end{cases}
\nonumber\\
\kappa &= \kappa^{\RN}=
\begin{cases}
(N-1)(N-2), & \RN=\AN, \cr
2N(N-1), & \RN=\BN, \CN, \cr
2(N-1)(N+1), & \RN=\BNv, \cr
(N-1)(2N-1), & \RN=\CNv, \cr
2N(N+1), & \RN=\BCN, \cr
2N(N-2), & \RN=\DN,
\end{cases}
\nonumber\\
g(\tau) &= g^{\RN}(\tau)=
\begin{cases}
1, & \RN=\AN, \BN, \CN, \DN, \cr
\displaystyle{
\left( \frac{\eta(2 \tau)}{\eta(\tau)^2} \right)^{2(N-1)} },
& \RN=\BNv, \cr
\displaystyle{
\left( \frac{\eta(\tau/2)^2}{\eta(\tau)} \right)^{N-1} },
& \RN=\CNv, \cr
\displaystyle{
\left( \frac{\eta(2 \tau)}{\eta(\tau)^2} \right)^{2N} },
& \RN=\BCN. \cr
\end{cases}
\label{eqn:factors1}
\end{align}
\end{lem}
\vskip 0.3cm
\noindent{\it Proof} \,
By the Macdonald denominator formulas (\ref{eqn:Macdonald3})
with (\ref{eqn:a1}),
\begin{align*}
Q^{\RN}(\z)
&= \frac{1}{a(\tau)^2}
\exp \left( - \frac{2 \pi \cN}{LW} \sum_{j=1}^N y_j^2 \right)
\det_{1 \leq j, k \leq N}
\Big[ \overline{M^{\RN}_j(z_k)} \Big]
\det_{1 \leq \ell, m \leq N}
\Big[ M^{\RN}_{\ell}(z_m) \Big]
\nonumber\\
&=  \frac{1}{a(\tau)^2}
\det_{1 \leq j, k \leq N}
\Big[e^{-\pi \cN y^2_k/(LW)} \overline{M^{\RN}_j(z_k)} \Big]
\det_{1 \leq \ell, m \leq N}
\Big[ e^{-\pi \cN y^2_m/(LW)} M^{\RN}_{\ell}(z_m) \Big].
\end{align*}
Use of the Heine identity (see, for instance, Eq.(C.4) in \cite{Kat18}) gives
\begin{align*}
& \frac{1}{N!} \int_{{\Lambda_{(L, iW)}}^N} 
\det_{1 \leq j, k \leq N}
\Big[e^{-\pi \cN y^2_k/(LW)} \overline{M^{\RN}_j(z_k)} \Big]
\det_{1 \leq \ell, m \leq N}
\Big[ e^{-\pi \cN y^2_m/(LW)} M^{\RN}_{\ell}(z_m) \Big] d \z
\nonumber\\
& \quad 
= \det_{1 \leq j, k \leq N}
\left[ \int_0^L dx \int_0^W dy \,
e^{-2 \pi \cN y^2/(LW)} 
\overline{M^{\RN}_j(z)} M^{\RN}_k(z) \right].
\end{align*}
Then Lemma \ref{thm:orthogonality_z} implies that
this multiple integral is given by
$\prod_{n=1}^N h^{\RN}_n$ and hence
\[
\frac{1}{N!}
\int_{{\Lambda_{(L, iW)}}^N} Q^{\RN}(\z) d \z
=\frac{1}{a(\tau)^2}
\prod_{n=1}^N h^{\RN}_n.
\]
From the expressions (\ref{eqn:ortho_const_z})
for $h^{\RN}_j, j \in \{1,2, \dots, N\}$,
the formulas (\ref{eqn:Qint1})
with (\ref{eqn:Z1}) and (\ref{eqn:factors1}) 
are obtained. \qed
\vskip 0.3cm

Let $\mM$ be the space of nonnegative 
integer-valued Radon measures on $\C$,
which is a Polish space with the vague topology:
we say $\xi_n, n \in \N \equiv \{1,2, \dots\}$ 
converges to $\xi$ vaguely, if 
$\lim_{n \to \infty} \int_{\C} \varphi(z) \xi_n(dz)
=\int_{\C} \varphi(z) \xi(dz)$ 
for any $\varphi \in {\rm C}_0$,
where ${\rm C}_0$ is the set of all 
continuous functions with
compact supports.
Any element $\xi$ of $\mM$ can be represented as
$\xi(\cdot) = \sum_{j\in \I}\delta_{z_j}(\cdot)$
with an index set $\I$, where
a sequence of points in $\C$, $\z =(z_j)_{j \in \I}$ 
satisfying $\xi(\cD)=\sharp\{j \in \I: z_j \in \cD \} < \infty$ 
for any compact domain $\cD \subset \C$.
Now we define seven types of point processes in $\mM$, 
\[
\Xi^{\RN}(\cdot)
=\sum_{j=1}^N \delta_{Z^{\RN}_j}(\cdot),
\]
for $\RN=\AN$, $\BN$, $\BNv$, $\CN$, $\CNv$, $\BCN$, $\DN$,
provided that
the number of particles in $\Lambda_{(L, iW)}$ is $N$
and the distributions of points
$\bZ^{\RN}=\{Z^{\RN}_j\}_{j=1}^N$ on $\C$ are
governed by the probability measures
\begin{equation}
\bP^{\RN}(\bZ^{\RN} \in d \z)
=\bp^{\RN}(\z) d\z
\equiv \frac{Q^{\RN}(\z)}{Z^{\RN}} d \z,
\label{eqn:P1}
\end{equation}
where 
\[
\int_{{\Lambda_{(L, iW)}}^N} \bp^{\RN}(\z) d \z
=1.
\]
The functions, 
$\bp^{\RN}(\z)$,
$\RN=\AN$, $\BN$, $\BNv$, $\CN$, $\CNv$, $\BCN$, $\DN$,
are called {\it probability density functions}.
By this construction, 
the probability density functions
are symmetric and doubly periodic
with periods $(L, iW)$ with respect to
$\z=(z_1, \dots, z_N) \in \C^N$.

Given the determinantal expressions
(\ref{eqn:QRN1}) and (\ref{eqn:P1}) 
with (\ref{eqn:Macdonald3}) for the
probability measures $\bP^{\RN}$ associated with
the orthogonal functions (Proposition \ref{thm:orthogonality_z}), 
we can readily prove the following 
by the standard method 
in random matrix theory \cite{Meh04,For10,AGZ10,Kat15_Springer}. 
(See, for instance, Appendix C in \cite{Kat18}.)

\begin{thm}
\label{thm:mainA1}
The seven types of point processes on $\C$, 
$(\Xi^{\RN}, \bP^{\RN})$,
$\RN=\AN$, $\BN$, $\BNv$, $\CN$, $\CNv$, $\BCN$, $\DN$, 
are determinantal with the correlation kernels, 
\begin{equation}
K^{\RN}(z, z^{\prime})  = 
e^{- \pi \cN \{y^2+(y')^2\}/(LW)}
\sum_{n=1}^{N}
\frac{1}{h^{\RN}_n} 
M^{\RN}_n(z) 
\overline{M^{\RN}_n(z^{\prime})}, 
\quad z, z^{\prime} \in \C.
\label{eqn:K1}
\end{equation}
These point processes are doubly periodic with periods
$(L, iW)$.
\end{thm}
\vskip 0.3cm
\noindent{\bf Remark 1} \,
The correlation kernels are quasi doubly periodic as shown below,
\begin{align*}
K^{\RN}(z+L, z^{\prime}) 
&= K^{\RN}(z, z^{\prime}+L)
\nonumber\\
&=\begin{cases}
(-1)^{\cN} K^{\RN}(z, z^{\prime}),
\quad & \RN=\AN,
\cr
- K^{\RN}(z, z^{\prime}),
\quad & \RN=\BN, \CNv, \BCN,
\cr
K^{\RN}(z, z^{\prime}),
\quad & \RN=\BNv, \CN, \DN,
\end{cases}
\nonumber\\
K^{\RN}(z+i W, z^{\prime})
&= \begin{cases}
e^{-2\pi i \cN x/L} K^{\RN}(z, z^{\prime}),
\quad & \RN=\AN, \CN, \CNv, \BCN, \DN,
\cr
-e^{-2\pi i \cN x/L} K^{\RN}(z, z^{\prime}),
\quad & \RN=\BN, \BNv,
\end{cases}
\nonumber\\
K^{\RN}(z, z^{\prime}+iW)
&= \begin{cases}
e^{2\pi i \cN x^{\prime}/L} K^{\RN}(z, z^{\prime}),
\quad & \RN=\AN, \CN, \CNv, \BCN, \DN,
\cr
-e^{2\pi i \cN x^{\prime}/L} K^{\RN}(z, z^{\prime}),
\quad & \RN=\BN, \BNv.
\end{cases}
\end{align*}
Since Theorem \ref{thm:mainA1} states that the present point 
processes are determinantal, 
for any $1 \leq N^{\prime} \leq N$,
$N^{\prime}$-point correlation function
at configuration 
$\z_{N^{\prime}}=(z_1, \dots, z_{N^{\prime}}) \in \C^{N^{\prime}}$
is given by
\begin{equation}
\rho^{\RN}(\z_{N^{\prime}})
= \det_{1 \leq j, k \leq N^{\prime}}
[K^{\RN}(z_j, z_k)].
\label{eqn:correlations}
\end{equation}
Although $K^{\RN}(z, z^{\prime})$ are quasi doubly periodic,
by basic property of determinant, 
the correlation functions (\ref{eqn:correlations}) 
are all doubly periodic with periods $(L, iW)$
with respect to $\z_{N^{\prime}}=(z_1, \dots, z_{N^{\prime}})$.
This is a matter of course, since the probability density functions
$\bp^{\RN}(\z)$ are symmetric and doubly periodic with respect to
$\z=(z_1, \dots, z_N)$, and the correlation functions are defined by
\[
\rho^{\RN}(\z_{N^{\prime}})
\equiv \frac{1}{(N-N^{\prime})!}
\int_{ {\Lambda_{(L, iW)}}^{N-N^{\prime}}}
\prod_{j=N^{\prime}}^{N} dz_j
\, \bp^{\RN}(z_1, \dots, z_{N^{\prime}},
z_{N^{\prime}+1}, \dots, z_{N}),
\quad \z_{N^{\prime}} \in \C^{N^{\prime}}.
\]

\subsection{DPPs with an infinite number of points}
\label{sec:infinite}

Now we consider the double limit $N \to \infty, L \to \infty$. 
We fix the density of points,
\begin{equation}
\rho=\frac{N}{LW}.
\label{eqn:rho1}
\end{equation}
and here we assume that the value of $W$ is also fixed.
This implies $\tau=i W/L= i \rho W^2/N \to 0$ in the limit $N \to \infty$.
Then we obtain the following.

\begin{prop}
\label{thm:infinite_systems1}
The following scaling limits are obtained for
correlation kernels. \\
{\rm (i)} For $\RN=\AN$,
\begin{align}
\cK^{A}_{W, \rho}(z, z^{\prime})
&\equiv \lim_{\substack{N \to \infty, L \to \infty, \cr
N/L=\rho W}}
K^{\AN}(z, z^{\prime})
\nonumber\\
&=\sqrt{2} \rho e^{-\pi \rho \{y^2+(y^{\prime})^2\}}
\int_0^{\sqrt{\rho} W}
d \lambda \,
e^{-2 \pi \lambda^2 + 2 \pi i \sqrt{\rho} (z-\overline{z^{\prime}}) \lambda}
\nonumber\\
& \qquad \qquad \times
\vartheta_2 ( \sqrt{\rho} W(i \lambda+\sqrt{\rho} z) ; i \rho W^2)
\vartheta_2 ( \sqrt{\rho} W(i \lambda-\sqrt{\rho} \overline{z^{\prime}}) ; 
i \rho W^2), 
\label{eqn:cKA}
\end{align}
$z, z^{\prime} \in \C$. \\
\noindent{\rm (ii)} For $\RN=\BN, \BNv$,
\begin{align}
& \cK^{B}_{W, \rho}(z, z^{\prime})
\equiv \lim_{\substack{N \to \infty, L \to \infty, \cr
N/L=\rho W}}
K^{\RN}(z, z^{\prime})
\nonumber\\
& \quad = - \rho e^{-2 \pi \rho \{y^2+(y^{\prime})^2\}}
\nonumber\\
& \qquad \times \left[
\int_{-\sqrt{\rho} W}^{\sqrt{\rho} W} d \lambda \,
e^{-\pi \lambda^2+2 \pi i \sqrt{\rho}(z-\overline{z^{\prime}}) \lambda}
\vartheta_1(\sqrt{\rho} W (i \lambda+2 \sqrt{\rho} z); 2 i \rho W^2)
\vartheta_1(\sqrt{\rho} W (i \lambda-2 \sqrt{\rho} \overline{z^{\prime}}); 
2 i \rho W^2)
\right.
\nonumber\\
& \qquad \left.
- \int_{-\sqrt{\rho} W}^{\sqrt{\rho} W} d \lambda \,
e^{-\pi \lambda^2+2 \pi i \sqrt{\rho}(z+\overline{z^{\prime}}) \lambda}
\vartheta_1(\sqrt{\rho} W (i \lambda+2 \sqrt{\rho} z); 2 i \rho W^2)
\vartheta_1(\sqrt{\rho} W (i \lambda+2 \sqrt{\rho} \overline{z^{\prime}}); 
2 i \rho W^2) \right],
\label{eqn:cKB}
\end{align}
$z, z^{\prime} \in \C$. \\
\noindent{\rm (iii)} For $\RN=\CN, \CNv, \BCN$,
\begin{align}
&\cK^{C}_{W, \rho}(z, z^{\prime})
\equiv \lim_{\substack{N \to \infty, L \to \infty, \cr
N/L=\rho W}}
K^{\RN}(z, z^{\prime})
\nonumber\\
& \quad = \rho e^{-2 \pi \rho \{y^2+(y^{\prime})^2\}}
\nonumber\\
& \qquad \times \left[
\int_{-\sqrt{\rho} W}^{\sqrt{\rho} W} d \lambda \,
e^{-\pi \lambda^2+2 \pi i \sqrt{\rho}(z-\overline{z^{\prime}}) \lambda}
\vartheta_2(\sqrt{\rho} W (i \lambda+2 \sqrt{\rho} z); 2 i \rho W^2)
\vartheta_2(\sqrt{\rho} W (i \lambda-2 \sqrt{\rho} \overline{z^{\prime}}); 
2 i \rho W^2)
\right.
\nonumber\\
& \qquad \left.
- \int_{-\sqrt{\rho} W}^{\sqrt{\rho} W} d \lambda \,
e^{-\pi \lambda^2+2 \pi i \sqrt{\rho}(z+\overline{z^{\prime}}) \lambda}
\vartheta_2(\sqrt{\rho} W (i \lambda+2 \sqrt{\rho} z); 2 i \rho W^2)
\vartheta_2(\sqrt{\rho} W (i \lambda+2 \sqrt{\rho} \overline{z^{\prime}}); 
2 i \rho W^2) \right],
\label{eqn:cKC}
\end{align}
$z, z^{\prime} \in \C$. \\
\noindent{\rm (iv)} For $\RN=\DN$,
\begin{align}
&\cK^{D}_{W, \rho}(z, z^{\prime})
\equiv \lim_{\substack{N \to \infty, L \to \infty, \cr
N/L=\rho W}}
K^{\DN}(z, z^{\prime})
\nonumber\\
& \, = \rho e^{-2 \pi \rho \{y^2+(y^{\prime})^2\}}
\nonumber\\
& \qquad \times \left[
\int_{-\sqrt{\rho} W}^{\sqrt{\rho} W} d \lambda \,
e^{-\pi \lambda^2+2 \pi i \sqrt{\rho}(z-\overline{z^{\prime}}) \lambda}
\vartheta_2(\sqrt{\rho} W (i \lambda+2 \sqrt{\rho} z); 2 i \rho W^2)
\vartheta_2(\sqrt{\rho} W (i \lambda-2 \sqrt{\rho} \overline{z^{\prime}}); 
2 i \rho W^2)
\right.
\nonumber\\
& \quad \left.
+ \int_{-\sqrt{\rho} W}^{\sqrt{\rho} W} d \lambda \,
e^{-\pi \lambda^2+2 \pi i \sqrt{\rho}(z+\overline{z^{\prime}}) \lambda}
\vartheta_2(\sqrt{\rho} W (i \lambda+2 \sqrt{\rho} z); 2 i \rho W^2)
\vartheta_2(\sqrt{\rho} W (i \lambda+2 \sqrt{\rho} \overline{z^{\prime}}); 
2 i \rho W^2) \right],
\label{eqn:cKD}
\end{align}
$z, z^{\prime} \in \C$. 
\end{prop}
\noindent{\it Proof} \, 
We fix the values of $\rho$ and $W$ and change the value of $N$. 
We have $\tau=i \rho W^2/N$ and $L=N/(\rho W)$. \\
First we consider the case (i).
$K^{\AN}(z,z^{\prime})$ is written as
\begin{align*}
K^{\AN}(z, z^{\prime})
&= \sqrt{2} \rho^{3/2} W e^{-\pi \rho \{y^2+(y^{\prime})^2\}}
\frac{1}{N} \sum_{n=1}^N e^{-2 \pi \rho W^2 (n-1/2)^2/N^2}
e^{2 \pi i \rho W (z-\overline{z^{\prime}}) (n-1/2)/N}
\nonumber\\
& \quad
\times \vartheta_2(i \rho W^2 (n-1/2)/N+\rho W z; i \rho W^2)
\vartheta_2(i \rho W^2 (n-1/2)/N-\rho W \overline{z^{\prime}}; i \rho W^2).
\end{align*}
This converges into the following integral in $N \to \infty$,
\begin{align*}
& \sqrt{2} \rho^{3/2} W e^{-\pi \rho\{y^2+(y^{\prime})^2\}}
\nonumber\\
& \quad \times
\int_0^1 d u \,
e^{-2 \pi \rho W^2 u^2+2 \pi i \rho W u (z-\overline{z^{\prime}})}
\vartheta_2(i \rho W^2 u+\rho W z; i \rho W^2)
\vartheta_2(i \rho W^2 u-\rho W \overline{z^{\prime}}; i \rho W^2).
\end{align*}
We change the integral valuable as $u \to \lambda=\sqrt{\rho} W u$,
and then (\ref{eqn:cKA}) is obtained. \\
Next we consider the case (iii). Using $\rho$ and $W$, 
$K^{\RN}(z, z^{\prime})$ is written as
\begin{align*}
K^{\RN}(z, z^{\prime})
&= \sqrt{\rho W^2 \frac{\cN}{2N}} \rho
e^{- 2 \pi \rho \{y^2+(y^{\prime})^2\} \cN/(2N)}
\nonumber\\
& \quad \times \frac{1}{N} \sum_{n=1}^N 
e^{-2 \pi \rho J(n)^2/(N \cN)}
\left[ e^{2 \pi i \rho W z J(n)/N}
\vartheta_2 \left(i \rho W^2 \frac{J(n)}{N} 
+  2 \rho W z \frac{\cN}{2N} \right) \right.
\nonumber\\
& \hskip 5cm \left.
- e^{-2 \pi i \rho W z J(n)/N}
\vartheta_2 \left(i \rho W^2 \frac{J(n)}{N} 
-  2 \rho W z \frac{\cN}{2N} \right) \right]
\nonumber\\
& \hskip 4cm \times
\left[ e^{-2 \pi i \rho W \overline{z^{\prime}} J(n)/N}
\vartheta_2 \left(i \rho W^2 \frac{J(n)}{N} 
-  2 \rho W \overline{z^{\prime}} \frac{\cN}{2N} \right) \right.
\nonumber\\
& \hskip 5cm \left.
- e^{2 \pi i \rho W \overline{z^{\prime}} J(n)/N}
\vartheta_2 \left(i \rho W^2 \frac{J(n)}{N} 
+  2 \rho W \overline{z^{\prime}} \frac{\cN}{2N} \right) \right].
\end{align*}
Since $\cN/(2N) \to 1$ as $N \to \infty$ for $\RN=\CN, \CNv, \BCN$,
this converges to the following integral in $N \to \infty$,
\begin{align*}
& \rho^{3/2} W e^{- 2 \pi \rho \{y^2+(y^{\prime})^2\}}
\nonumber\\
& \quad \times \int_0^1 du \, e^{-\pi \rho W^2 u^2}
\Big[ e^{2 \pi i \rho W z u}
\vartheta_2(i \rho W^2 u + 2 \rho W z; 2 i \rho W^2)
-  e^{- 2 \pi i \rho W z u}
\vartheta_2(i \rho W^2 u - 2 \rho W z; 2 i \rho W^2) \Big]
\nonumber\\
& \hskip 2.5cm \times
\Big[ e^{-2 \pi i \rho W \overline{z^{\prime}} u}
\vartheta_2(i \rho W^2 u - 2 \rho W \overline{z^{\prime}}; 2 i \rho W^2)
-  e^{2 \pi i \rho W \overline{z^{\prime}} u}
\vartheta_2(i \rho W^2 u + 2 \rho W \overline{z^{\prime}}; 2 i \rho W^2) \Big].
\end{align*}
If we change the integral variable as $u \to \lambda=\sqrt{\rho} W u$
and use the fact that $\vartheta_2(-v; \tau)=\vartheta_2(v; \tau)$,
(\ref{eqn:cKC}) is obtained. \\
The cases (ii) and (iv) are proved by the similar calculation.
The proof is complete. \qed
\vskip 0.3cm

The limit correlation kernels have the following quasi periodicity
with period $i W$,
\begin{align*}
\cK^{R}_{W, \rho}(z+i W, z^{\prime})
&= \begin{cases}
e^{-2\pi i \rho W x} \cK^{R}_{W, \rho}(z, z^{\prime}),
\quad & R=A, C, D,
\cr
-e^{-2\pi i \rho W x} \cK^{R}_{W, \rho}(z, z^{\prime}),
\quad & R=B, 
\end{cases}
\nonumber\\
\cK^{R}_{W, \rho}(z, z^{\prime}+iW)
&= \begin{cases}
e^{2\pi i \rho W x^{\prime}} \cK^{R}_{W, \rho}(z, z^{\prime}),
\quad & R=A, C, D,
\cr
-e^{2\pi i \rho W x^{\prime}} \cK^{R}_{W, \rho}(z, z^{\prime}),
\quad & R=B. 
\end{cases}
\end{align*}

We see that the limit kernels, $\cK^{R}_{W, \rho}(z, z^{\prime})$,
$R=A, B, C, D$, 
are continuous functions of $(z, z^{\prime}) \in \C^2$
and
\[
\sup_{z, z^{\prime} \in \cD} |K^{\RN}(z, z^{\prime})|
< \infty, \quad \forall N \in \N,
\]
for any compact domain $\cD \subset \C$. 
Then we can obtain the convergence of generating functions
of correlation functions, which are expressed by
Fredholm determinants \cite{KT10,Kat15_Springer}, as $N \to \infty$.
Generating functions of correlation functions can be
identified with the Laplace transforms of probability densities of 
point processes.
Hence this implies the convergence of probability laws weakly
in the sense of finite dimensional distributions in the
vague topology \cite{KT10,Kat15_Springer}.
We write the probability laws for the infinite dimensional
DPPs associated with the
limit kernels $\cK^{R}_{W, \rho}$, $R=A, B, C, D$
given in Proposition \ref{thm:infinite_systems1} as
$\cP^{R}_{W, \rho}$, $R=A, B, C, D$, respectively. 

The density profile of DPP is given by 
$\rho^{R}_{W, \rho}(z)=\cK^{R}_{W, \rho}(z,z), z \in \C$.
It is easy to verify that 
$\cK^{B}_{W, \rho}(0, 0)=\cK^{C}_{W, \rho}(0, 0)=0$
for (\ref{eqn:cKB}) and (\ref{eqn:cKC}). 

\begin{thm}
\label{thm:scaling_limits}
In the scaling limit $N \to \infty$, $L \to \infty$
with constant density of points {\rm (\ref{eqn:rho1})}
and constant $W$, 
the seven types of elliptic DPPs on $\C$, 
$(\Xi^{\RN}, \bP^{\RN})$,
$\RN=\AN$, $\BN$, $\BNv$, $\CN$, $\CNv$, $\BCN$, $\DN$,
converge weakly in the sense of finite dimensional distributions
in the vague topology to
the four types of infinite dimensional DPPs
as follows, 
\begin{align*}
(\Xi^{\AN}, \bP^{\AN})
&\Longrightarrow
(\Xi^{A}, \cP^{A}_{W, \rho}),
\nonumber\\
\left.
\begin{array}{l}
(\Xi^{\BN}, \bP^{\BN}) \cr
(\Xi^{\BNv}, \bP^{\BNv}) 
\end{array}
\right\}
&\Longrightarrow
(\Xi^{B}, \cP^{B}_{W, \rho}), 
\nonumber\\
\left.
\begin{array}{l}
(\Xi^{\CN}, \bP^{\CN}) \cr
(\Xi^{\CNv}, \bP^{\CNv}) \cr
(\Xi^{\BCN}, \bP^{\BCN}) 
\end{array}
\right\}
&\Longrightarrow
(\Xi^{C}, \cP^{C}_{W, \rho}), 
\nonumber\\
(\Xi^{\DN}, \bP^{\DN})
&\Longrightarrow
(\Xi^{D}, \cP^{D}_{W, \rho}).
\end{align*}
The four types of infinite DPPs on $\C$,
$(\Xi^{R}, \cP^{R}_{W, \rho}), R=A, B, C, D$, 
have the periodicity with period $i W$.
In particular, the densities
at the points $\{in W\}_{n \in \Z}$ on the imaginary axis
are fixed to be zero in $\cP^{B}_{W, \rho}$ and $\cP^{C}_{W, \rho}$;
\[
\rho^{B}_{W, \rho}(i n W)=\rho^{C}_{W, \rho}(i n W)=0,
\quad n \in \Z.
\]
\end{thm}
\vskip 0.3cm

\noindent{\bf Remark 2} \,
Using (\ref{eqn:Theta}), we define
\begin{align*}
g^{A}_{W, \rho}(z, \lambda)
&= 2^{1/4} \sqrt{\rho} e^{-\pi(\rho y^2+\lambda^2)}
\Theta^A \left( \frac{\lambda}{\sqrt{\rho} W}, \rho W z, i \rho W^2 \right),
\nonumber\\
g^{R}_{W, \rho}(z, \lambda)
&= \sqrt{\rho} e^{-\pi(2 \rho y^2+\lambda^2/2)}
\Theta^R \left( \frac{\lambda}{2 \sqrt{\rho} W}, 2 \rho W z, 2 i \rho W^2 \right),
\quad R=B, C, D.
\end{align*}
We consider a strip in $\C$ given by
\[
\cD_W = \{ z \in \C : 0 \leq \Im z < W \} = \R \times i[0, W),
\quad W>0.
\]
Then provided $\lambda, \lambda' \in (0, \sqrt{\rho} W)$, $\rho >0$,
we can prove the following orthonormality relations,
\[
\int_{\cD_W} 
\overline{g^{R}_{W, \rho}(z, \lambda)} g^{R}_{W, \rho}(z, \lambda') d z
=\delta(\lambda-\lambda'),
\quad R=A, B, C, D.
\]
It is easy to verify that the four kernels in Proposition \ref{thm:infinite_systems1}
are written as
\begin{equation}
\cK^{R}_{\rho, W}(z, z')
=\int_0^{\sqrt{\rho} W} g^{R}_{W, \rho}(z, \lambda)
\overline{g^{R}_{W, \rho}(z', \lambda)} d \lambda,
\quad z, z' \in \C, \quad
R=A, B, C, D.
\label{eqn:projection}
\end{equation}
Hence the four kernels obtained in the scaling limit
$N \to \infty, L \to \infty$ with fixed 
$\rho=N/(LW)$ and $W$ are all {\it reproducing kernels}.
Since (\ref{eqn:projection}) implies that
$\cK^{R}_{\rho, W}$, $R=A, B, C, D$, are regarded as
the integral kernels giving {\it orthonormal projections},
we can conclude that they indeed provide correlation kernels
of DPPs \cite{Sos00,ST03,ST03b}.
More detail, see \cite{KS}.
\vskip 0.3cm
\noindent{\bf Remark 3} \,
We find a resemblance of $\cK^{R}_{W, \rho}$ in 
Proposition \ref{thm:infinite_systems1} to the correlation kernel
$K_{\rm weak}$ obtained in the
{\it limit of weak non-Hermiticity} \cite{FKS98,FS03}
for the eigenvalue distribution of a non-Gaussian extension
of the elliptic Ginibre ensemble of complex non-Hermitian
random matrices.
It is given in Theorem 3 (b) of \cite{ACV18} as
\[
K_{\rm weak}(z, z^{\prime})
=\frac{\sqrt{2}}{\sqrt{\pi} \alpha}
e^{-\{y^2+(y^{\prime})^2\}/\alpha^2}
\frac{1}{2 \pi}
\int_{-\pi}^{\pi} e^{-\alpha^2 u^2/2+iu(z-\overline{z^{\prime}})} du, 
\]
with a parameter $\alpha>0$. 
Actually, if we set $\alpha=1/\sqrt{\pi \rho}$ and
change the integral variable $\lambda \to u$ by
$\lambda=\alpha u/(2 \sqrt{\pi})$,
$\cK^{A}_{W, \rho}$ given by (\ref{eqn:cKA}) is written as
\begin{align*}
\cK^A_{W, \rho}(z, z^{\prime})
&= 
\frac{\sqrt{2}}{\sqrt{\pi} \alpha}
e^{-\{y^2+(y^{\prime})^2\}/\alpha^2}
\frac{1}{2 \pi}
\int_{0}^{W/(\sqrt{\pi} \alpha)} e^{-\alpha^2 u^2/2+iu(z-\overline{z^{\prime}})} 
\nonumber\\
& \hskip 3cm \times
\vartheta_2 \left( \frac{W}{2 \pi} \left( iu + \frac{2z}{\alpha^2} \right); 
\frac{i W^2}{\pi \alpha^2} \right)
\vartheta_2 \left( \frac{W}{2 \pi} \left( iu - \frac{2z}{\alpha^2} \right); 
\frac{i W^2}{\pi \alpha^2} \right) du.
\end{align*}
Since our correlation kernels $\cK^{R}_{W, \rho}$ contain
the Jacobi theta functions, they will be considered as
elliptic extensions of $K_{\rm weak}$ of
\cite{FKS98,FS03,ACV18}.
The {\it limit of strong non-Hermiticity} associated with
$\alpha \to \infty$ discussed in Remark 4 (c) in \cite{ACV18}
will be achieved in the limit $W \to \infty$ in DPPs
as shown below.
See also \cite{Osb04}, Section 15.11 in \cite{For10}, 
and references therein
for non-Hermitian random-matrix models. 
We do not know, however, as eigenvalue distributions
what kinds of random matrix ensembles
realize the present elliptic DPPs on $\C$,
$(\Xi^R, \cP^R_{W, \rho}), R=A, B, C, D$.
\vskip 0.3cm

Now we consider the further limit
\[
W \to \infty \quad \mbox{with constant $\rho$}.
\]
In this limit, $\Im(i \rho W^2)=\rho W^2 \to \infty$,
and application of (\ref{eqn:theta_asym}) to (\ref{eqn:cKA}) gives
\begin{align*}
\cK^{A}_{W, \rho}(z, z^{\prime})
&\simeq 4 \sqrt{2} \rho e^{-\pi \rho \{y^2+(y^{\prime})^2\}}
\int_0^{\sqrt{\rho} W}
d \lambda \,
e^{-2 \pi \lambda^2 + 2 \pi i \sqrt{\rho}(z-\overline{z^{\prime}}) \lambda
-\pi \rho W^2/2}
\nonumber\\
& \qquad \qquad
\times
\cos[ \pi \sqrt{\rho} W(i\lambda+\sqrt{\rho}z) ]
\cos[ \pi \sqrt{\rho} W(i\lambda-\sqrt{\rho} \overline{z^{\prime}}) ]
\quad \mbox{ as $W \to \infty$}.
\end{align*}
We can rewrite this as
\begin{align*}
\cK^{A}_{W, \rho}(z, z^{\prime})
&\simeq \sqrt{2} \rho e^{-\pi \rho \{ y^2+(y^{\prime})^2\}} 
\left[
e^{-\pi \rho(z-\overline{z^{\prime}})^2/2}
\int_{-\sqrt{\rho}\{W-(y+y^{\prime})\}/2}
^{\sqrt{\rho}\{3W+(y+y^{\prime})\}/2}
e^{-2 \pi \eta^2} d \eta \right.
\nonumber\\
& \qquad \qquad \left.
+ 2 e^{-\pi \rho W^2/2-\pi \rho(z-\overline{z^{\prime}})^2/2}
\cos [ \pi \rho W(z+\overline{z^{\prime}})]
\int_{\sqrt{\rho}(y+y^{\prime})/2}
^{\sqrt{\rho}\{2 W+(y+y^{\prime})\}/2}
e^{-2 \pi \eta^2} d \eta
\right]
\end{align*}
as $W \to \infty$. 
Then we obtain the limit
\begin{align*}
\lim_{W \to \infty} \cK^{A}_{W, \rho}(z,z^{\prime})
&= \rho e^{-\pi \rho \{y^2+(y^{\prime})^2\}}
e^{-\pi \rho (z-\overline{z^{\prime}})^2/2}
\nonumber\\
&= \frac{e^{\pi i \rho x^{\prime} y^{\prime}}}{e^{\pi i \rho x y}}
\rho e^{-\pi \rho(|z|^2+|z^{\prime}|^2)/2 + \pi \rho z \overline{z^{\prime}}}.
\end{align*}
Since any factor of the form $f(z^{\prime})/f(z)$ is irrelevant
in determinantal correlation kernels, 
by basic property of determinant, we can identify the limit kernel with
\begin{equation}
\cK^{A}_{{\rm Ginibre}, \rho}(z,z^{\prime})
\equiv \rho e^{-\pi \rho(|z|^2+|z^{\prime}|^2)/2 + \pi \rho z \overline{z^{\prime}}},
\label{eqn:GinibreA}
\end{equation}
which is the correlation kernel for the
{\it Ginibre ensemble} 
studied in random matrix theory \cite{Gin65,HKPV09,Shi15}
with uniform density
\begin{equation}
\rho^{A}_{{\rm Ginibre}, \rho}(z)
=\cK^{A}_{{\rm Ginibre}, \rho}(z, z)=\rho,
\quad z \in \C.
\label{eqn:rho_G_A}
\end{equation}

Similarly, we can obtain the following limit kernels, 
\begin{align*}
\lim_{W \to \infty} \cK^{B}_{W, \rho}(z, z^{\prime})
&= \lim_{W \to \infty} \cK^{C}_{W, \rho}(z, z^{\prime})
\nonumber\\
&= \rho e^{-2 \pi \rho \{y^2+(y^{\prime})^2\}}
\Big[ e^{-\pi \rho(z-\overline{z^{\prime}})^2}
-e^{-\pi \rho(z+\overline{z^{\prime}})^2} \Big]
\nonumber\\
&=\frac{e^{2 \pi i \rho x^{\prime} y^{\prime}}}{e^{2 \pi i \rho x y}}
\cK^{C}_{{\rm Ginibre}, \rho}(z,z^{\prime}),
\nonumber\\
\lim_{W \to \infty} \cK^{D}_{W, \rho}(z, z^{\prime})
&= \rho e^{-2 \pi \rho \{y^2+(y^{\prime})^2\}}
\Big[ e^{-\pi \rho(z-\overline{z^{\prime}})^2}
+e^{-\pi \rho(z+\overline{z^{\prime}})^2} \Big]
\nonumber\\
&=\frac{e^{2 \pi i \rho x^{\prime} y^{\prime}}}{e^{\pi i \rho x y}}
\cK^{D}_{{\rm Ginibre}, \rho}(z,z^{\prime}),
\end{align*}
where
\begin{align}
\cK^{C}_{{\rm Ginibre}, \rho}(z,z^{\prime})
&\equiv 2 \rho e^{-\pi \rho(|z|^2+|z^{\prime}|^2)}
\sinh( 2\pi \rho  z \overline{z^{\prime}} ), 
\label{eqn:GinibreC}
\\
\cK^{D}_{{\rm Ginibre}, \rho}(z,z^{\prime})
&\equiv 2 \rho e^{-\pi \rho(|z|^2+|z^{\prime}|^2)}
\cosh( 2\pi \rho  z \overline{z^{\prime}} ), 
\quad z \in \C.
\label{eqn:GinibreD}
\end{align}
We write the probability laws of the DPPs
governed by $\cK^{R}_{{\rm Ginibre}, \rho}(z,z^{\prime})$
as $\cP^{R}_{{\rm Ginibre}, \rho}$ for
$R=A, C$, and $D$.

\begin{thm}
\label{thm:Ginibre}
In the limit $W \to \infty$
with constant density of points $\rho$, 
the four types of DPPs on $\C$, 
$(\Xi^{R}, \cP^{R}_{W, \rho})$,
$R=A, B, C, D$
converge weakly in the sense of finite dimensional distributions
in the vague topology to
the three types of infinite dimensional point processes
as follows, 
\begin{align*}
(\Xi^{A}, \cP^{A}_{W, \rho})
&\Longrightarrow
(\Xi^{A}, \cP^{A}_{{\rm Ginibre}, \rho}),
\nonumber\\
\left.
\begin{array}{l}
(\Xi^{B}, \cP^{B}_{W, \rho}) \cr
(\Xi^{C}, \cP^{C}_{W, \rho}) 
\end{array}
\right\}
&\Longrightarrow
(\Xi^{C}, \cP^{C}_{{\rm Ginibre}, \rho}), 
\nonumber\\
(\Xi^{D}, \cP^{D}_{W, \rho})
&\Longrightarrow
(\Xi^{D}, \cP^{D}_{{\rm Ginibre}, \rho}).
\end{align*}
\end{thm}

From (\ref{eqn:GinibreC}) and (\ref{eqn:GinibreD}), we see that
the density profiles are rotationally symmetric around the origin, 
but not uniform on $\C$ in
$\cP^{R}_{{\rm Ginibre}, \rho}$ for $R=C$ and $D$;
\begin{align}
\cK^{C}_{{\rm Ginibre}, \rho}(z,z)
&= \rho^{C}_{{\rm Ginibre}, \rho}(z)
=\rho^{C}_{{\rm Ginibre}, \rho}(|z|) 
\nonumber\\
&= 2 \rho \sinh(2 \pi \rho |z|^2) e^{-2 \pi \rho |z|^2}
=\rho[1-e^{-4 \pi \rho |z|^2}],
\label{eqn:rho_G_C}
\\
\cK^{D}_{{\rm Ginibre}, \rho}(z,z)
&=\rho^{D}_{{\rm Ginibre}, \rho}(z)
=\rho^{D}_{{\rm Ginibre}, \rho}(|z|) 
\nonumber\\
&= 2 \rho \cosh(2 \pi \rho |z|^2) e^{-2 \pi \rho |z|^2}
= \rho[1+e^{-4 \pi \rho |z|^2}],
\label{eqn:rho_G_D}
\end{align}
in which 
\begin{align}
&\min_{z \in \C} \rho^{C}_{{\rm Ginibre}, \rho}(z)
=\rho^{C}_{{\rm Ginibre}, \rho}(0)=0,
\label{eqn:rho_C_0}
\\
&\max_{z \in \C} \rho^{D}_{{\rm Ginibre}, \rho}(z)
=\rho^{D}_{{\rm Ginibre}, \rho}(0)=2 \rho,
\label{eqn:rho_D_0}
\end{align}
and
\begin{equation}
\lim_{|z| \to \infty}
\rho^{R}_{{\rm Ginibre}, \rho}(z) = \rho,
\quad R=C, D.
\label{eqn:rho_inf_CD}
\end{equation}

\vskip 0.3cm
\noindent{\bf Remark 4} \,
Combination of Theorem \ref{thm:scaling_limits}
and Theorem \ref{thm:Ginibre} implies that
in the {\it bulk scaling limit},
\[
N \to \infty, \quad L \to \infty, \quad W \to \infty
\quad \mbox{ with constant
$\displaystyle{ \rho=\frac{N}{LW}}$ }, 
\]
the seven types of elliptic DPPs, 
$(\Xi^{\RN}, \bP^{\RN})$,
$\RN=\AN$, $\BN$, $\BNv$, $\CN$, $\CNv$, $\BCN$, $\DN$,
converge weakly in the sense of finite dimensional distributions
in the vague topology to
the three types of Ginibre-like DPPs,
$(\Xi^{R}, \cP^{R}_{{\rm Ginibre}, \rho})$,
$R=A, C, D$.
\vskip 0.3cm
\noindent{\bf Remark 5} \,
Recently a general theory has been developed
for the one-point functions (the density profiles)
of point processes in $\C$ in \cite{AK13,AKM18,AKS18}.
The one-mass theorem in \cite{AK13} can be interpreted
as the statement that, if $\I$ is an index set
given by a nonempty subset of 
$\N_0 \equiv \{0,1,2, \dots\}$, then
\begin{equation}
\rho^{\I}_{k, c}(z)
= k \sum_{j \in \I} \frac{|z|^{2j}}{\Gamma(j/k+(1+c)/k)}
e^{-Q_{k, c}(z)}, 
\label{eqn:ML1}
\end{equation}
with the microscopic potential
\[
Q_{k,c}(z)=|z|^{2k}-2c \log |z|,
\]
defines the one-point function of a unique
rotationally symmetric point process in $\C$.
Here $\Gamma(z)$ is the gamma function,
$k \in \N$, and $c \in \R$
represents an inserted point charge of strength $c$
at the origin \cite{AKS18}.
when $\I=\N_0$, (\ref{eqn:ML1}) is written as
\[
\rho^{\N_0}_{1,1}(z)
= k E_{1/k, (1+c)/k}(|z|^2) e^{-Q_{k,c}(z)},
\]
where $E_{a,b}(z)$ is 
the two-parametric Mittag--Leffler function \cite{GKMR14}
defined as
\[
E_{a,b}(z)=\sum_{j=0}^{\infty} \frac{z^j}{\Gamma(aj+b)}.
\]
For any $a, b \in \C, \Re a >0$, $E_{a, b}(z)$ is 
an entire function of order $1/(\Re \alpha)$ and type 1.
The rotationally symmetric point processes with an infinite
number of points in $\C$ is hence called
the {\it Mittag--Leffler fields} \cite{AKS18}.
The function $E_{a,b}(z)$ is a two-parameter extension of the
exponential $E_{1,1}(z)=e^z$, and when we choose
$k=1$ and $c=0$, we have 
$\rho^{\N_0}_{1, 0}(z) \equiv 1, z \in \C$.
It is easy to verify that, if we set $k=1, c=1$ and
$\I=2 \N_0 = \{0,2,4, \dots\}$, then (\ref{eqn:ML1}) gives
\[
\rho^{2\N_0}_{1,1}(z)=\sinh(|z|^2) e^{-|z|^2},
\]
and that, if we set $k=1, c=-1$ and $\I=2\N_0+1=\{1,3,5, \dots\}$,
then 
\[
\rho^{2\N_0+1}_{1,-1}(z)=\cosh(|z|^2) e^{-|z|^2}.
\]
The density profiles (\ref{eqn:rho_G_A}), (\ref{eqn:rho_G_C}),
and (\ref{eqn:rho_G_D}) are scale changes of
$\rho^{\N_0}_{1, 0}(z)$, $\rho^{2 \N_0}_{1,1}(z)$, and
$\rho^{2\N+1}_{1, -1}(z)$, respectively.
In this sense, the three types of Ginibre-like DPPs,
$(\Xi^{R}, \cP^R_{{\rm Ginibre}, \rho})$, $R=A, C, D$, given by
Theorem \ref{thm:Ginibre} can be regarded as new examples
of the Mittag--Leffler fields \cite{AK13,AKM18,AKS18}.
In Section \ref{sec:plasma_GFF}, we will construct two kinds of
exactly solvable one-component plasma models
from the elliptic DPPs of types $\CN$ and $\DN$
in addition to Forrester's model (of type $\AN$) \cite{For06}.
For our new models, we assume the two-point potential functions
between charged particles in the form (\ref{eqn:Phipm0}),
which is different from (\ref{eqn:Phi-0}) used by Forrester,
and our plasma model of type $\CN$ (resp. $\DN$)
is negatively (resp. positively) charged by unit,
while Forrester's model is perfectly neutralized.
These plasma models will be regarded as
`physical realizations' of the above mentioned settings of
index set $\I$ and parameter $c$ for the
new examples of the Mittag--Leffler fields.
See Remark 8 at the end of Section \ref{sec:solvability}.

\vskip 0.3cm
\noindent{\bf Remark 6} \,
Abreu {\it el al.} introduced a class of DPPs
called the {\it Weyl--Heisenberg ensembles}
associated with the Schr\"{o}dinger representation
of the Heisenberg group \cite{APRT17,AKR17}.
The Ginibre ensemble (of type $A$) with 
correlation kernel (\ref{eqn:GinibreA}) is included 
in this class as a typical example, but the DPPs
with the correlation kernels
$\cK^{R}_{{\rm Ginibre}, \rho}$, $R=C, D$,
given by (\ref{eqn:GinibreC}) and (\ref{eqn:GinibreD}) are not.
We have found, however, the Weyl--Heisenberg ensembles
can be extended to a wider class of DPPs \cite{KS}.
The following facts imply that all three types of Ginibre-like
DPPs, $(\Xi^{R}, \cP^{R}_{{\rm Ginibre}, \rho}), R=A, C, D$,
are included in our new class of DPP.
For $\rho>0, z \in \C, \lambda \in \R$, let
\begin{align*}
g^A_{\rho}(z, \lambda)
&= 2^{1/4} \rho^{3/4} e^{2 \pi \rho z \lambda - \pi \rho(x^2+\lambda^2)},
\nonumber\\
g^C_{\rho}(z, \lambda)
&= 2^{3/2} \rho^{3/4} \sinh(4 \pi \rho z \lambda)
e^{-2 \pi \rho (x^2+\lambda^2)},
\nonumber\\
g^D_{\rho}(z, \lambda)
&= 2^{3/2} \rho^{3/4} \cosh(4 \pi \rho z \lambda)
e^{-2 \pi \rho (x^2+\lambda^2)},
\end{align*}
and consider the space $\cS=\R$ for $R=A$ and
$\cS=(0, \infty)$ for $R=C, D$. 
Then the following orthonormality relations hold,
\begin{align*}
& \int_{\C} \overline{g^R_{\rho}(z, \lambda)}
g^R_{\rho}(z, \lambda^{\prime}) dz = \delta(\lambda-\lambda^{\prime}),
\quad \mbox{if $\lambda, \lambda^{\prime} \in \cS$},
\quad \mbox{for $R=A, C, D$}.
\end{align*}
And we can show that the correlation kernels
$\cK^{R}_{{\rm Ginibre}, \rho}$, $R=A, C, D$ given by
(\ref{eqn:GinibreA}), (\ref{eqn:GinibreC}),
and (\ref{eqn:GinibreD}),
are identified (up to irrelevant factors for
determinantal correlation kernels) with
the reproducing and orthonormal projection kernels
\[
\widetilde{\cK}^R_{\rho}(z, z^{\prime})
=\int_{\cS} g^{R}_{\rho}(z, \lambda) 
\overline{g^R_{\rho}(z^{\prime}, \lambda)} d \lambda,
\quad z, z^{\prime} \in \C,
\quad R=A, C, D,
\]
respectively.
More precise statements 
and another construction of the correlation kernels
$\cK^R_{{\rm Ginibre}, \rho}, R=A, C, D$, 
as projection kernels in the Bargmann--Fock space 
will be reported in \cite{KS}.

\SSC
{Realization as One-Component Plasma Systems
and Relationship to Gaussian Free Field} 
\label{sec:plasma_GFF}
\subsection{General setting of one-component plasma models}
\label{sec:plasma}

We consider the following two types of two-point 
potential functions in the rectangular domain
$\Lambda_{(L, i W)} \subset \C$;
\begin{align}
\Phi^{-}_{0}(z, z^{\prime})
&=\Phi^{-}_{0}(z, z^{\prime}; \tau)
\equiv - \log(|\vartheta_1((z-z^{\prime})/L); \tau)|),
\label{eqn:Phi-0}
\\
\Phi^{\pm}_{0}(z, z^{\prime})
&=\Phi^{\pm}_{0}(z, z^{\prime}; \tau)
\equiv - 
\log(|\vartheta_1( (z+z^{\prime})/L; \tau)| 
|\vartheta_1( (z-z^{\prime})/L; \tau)|)
\nonumber\\
&= - \Big\{
\log(|\vartheta_1( (z+z^{\prime})/L; \tau)|)
+\log(|\vartheta_1( (z-z^{\prime})/L; \tau)|)
\Big\},
\label{eqn:Phipm0}
\end{align}
$z, z^{\prime} \in \Lambda_{(L, i W)}$.

\vskip 0.3cm
\noindent{\bf Remark 7} \
Note that $\Phi^{\pm}_0(z, z^{\prime})$ is 
different from the two-point potential functions
in the system with the anti-metallic boundary condition
found in Section 15.9 in \cite{For10},
since, for $\xi=\Re \xi + i \Im \xi$,
$\overline{\xi}=\Re \xi- i \Im \xi
\not= - \xi = - \Re \xi- i \Im \xi$, in general.
With the potential $\Phi^{\pm}_0$, if a particle is at
$z \in \C$, it will repel both the position $z$ 
and the position $-z$ simultaneously.
For random matrix models with such interactions,
see \cite{Osb04}, Chapter 15.11 in \cite{For10},
and references therein.
\vskip 0.3cm

These two-point potential functions are related with the absolute values of the
complex-valued Macdonald denominators
of types $\AN$ and $\DN$ as
\begin{align*}
|W^{\AN}(\z/L; \tau)|
&= \exp \left( - \sum_{1 \leq j < k \leq N} \Phi^{-}_0(z_k, z_j; \tau) \right),
\nonumber\\
|W^{\DN}(\z/L; \tau)|
&= \exp \left( - \sum_{1 \leq j < k \leq N} \Phi^{\pm}_0(z_k, z_j; \tau) \right),
\quad \z =(z_1, z_2, \dots, z_N) \in {\Lambda_{(L, iW)}}^N.
\end{align*}

Since 
$\vartheta_1(\xi; \tau) \simeq \xi \partial \vartheta_1(\xi; \tau)/\partial \xi|_{\xi=0}$
as $\xi \to 0$, the relation (\ref{eqn:diff_theta1}) gives
\begin{align*}
\Phi^{-}_{0}(z, z^{\prime})
&\simeq - \log \left(
\eta(\tau)^3 \frac{2 \pi}{L} |z-z^{\prime}| \right),
\nonumber\\
\Phi^{\pm}_{0}(z, z^{\prime})
&\simeq - \log \left(
\eta(\tau)^3 \frac{2 \pi}{L} |z-z^{\prime}| \right)
-\frac{1}{2} \Big\{
\log(|\vartheta_1(2 z/L; \tau)|)
+ \log(|\vartheta_1(2 z^{\prime}/L; \tau)|) \Big\},
\end{align*}
as $|z-z^{\prime}| \to 0$.
Hence, if we define
\begin{align}
\Phi^{-}(z, z^{\prime})
&= \Phi^{-}_0(z, z^{\prime})
+ 3 \log(\eta(\tau))+\log(2\pi/L),
\label{eqn:Phi-}
\\
\Phi^{\pm}(z, z^{\prime})
&= \Phi^{\pm}_0(z, z^{\prime})
+ 3 \log(\eta(\tau))+\log(2\pi/L)
+ \frac{1}{2} \Big\{
\log(|\vartheta_1(2 z/L; \tau)|)
+ \log(|\vartheta_1(2 z^{\prime}/L; \tau)|) \Big\},
\label{eqn:Phipm}
\end{align}
then they have the common asymptotic form,
\[
\Phi^{-}(z, z^{\prime}) \simeq - \log|z-z^{\prime}|,
\quad
\Phi^{\pm}(z, z^{\prime}) \simeq - \log|z-z^{\prime}|,
\quad \mbox{as $|z-z^{\prime}| \to 0$},
\]
which solves the two-dimensional Poisson equation
\[
\left( \frac{\partial^2}{\partial x^2}
+\frac{\partial^2}{\partial y^2} \right)
u(x+iy, x^{\prime}+i y^{\prime})
=-2 \pi \delta(x-x^{\prime}) \delta(y-y^{\prime})
\]
with free boundary condition.

We consider two types of systems
of $N$ particles each of which has charge +1.
The $N$ particles are mobile and interacting
via the potential (\ref{eqn:Phi-}) or (\ref{eqn:Phipm}),
but they are confined to the rectangle domain
$\Lambda_{(L, i W)} \subset \C$.
The former system was studied by Forrester \cite{For06},
but the latter is new.
The total energies of the particle-particle interaction
in particle configurations $\z$ are given by
\begin{align*}
E^{-}_{\rm pp}(\z)
&= \sum_{1 \leq j, k \leq N} \Phi^{-}(z_k, z_j)
\nonumber\\
&= - \log \left(
\prod_{1 \leq j < k \leq N}
|\vartheta_1( (z_k-z_j)/L; \tau)| \right)
+ \frac{3}{2} N(N-1) \log(\eta(\tau))
+ \frac{1}{2} N(N-1) \log(2 \pi/L),
\nonumber\\
E^{\pm}_{\rm pp}(\z)
&= \sum_{1 \leq j, k \leq N} \Phi^{\pm}(z_k, z_j)
\nonumber\\
&= - \log \left(
\prod_{1 \leq j < k \leq N}
|\vartheta_1( (z_k-z_j)/L; \tau)| 
|\vartheta_1( (z_k+z_j)/L; \tau)| \right)
\nonumber\\
& \quad 
+ \frac{3}{2} N(N-1) \log(\eta(\tau))
+ \frac{1}{2} N(N-1) \log(2 \pi/L)
+\frac{N-1}{2} \log
\left( \prod_{j=1}^N |\vartheta_1(2 z_j/L; \tau)| \right),
\end{align*}
respectively.

We think that in both types of systems there are uniform 
backgrounds which are negatively charged and the negative charge
densities are given by $-N^{-}/LW$.
Here $N^{-}$ denotes a total number of negative charges
constructing the background, whose value is not equal to $N$
in general and will be determined later.
The particle-background potential energies are given by
\[
E^{\sharp}_{\rm pb}(\z)
= \sum_{j=1}^N V^{\sharp}(z_j),
\]
with
\begin{equation}
V^{\sharp}(z)
=- \frac{N^{-}}{LW} \int_0^L dx^{\prime}
\int_0^{W} dy^{\prime} \, \Phi^{\sharp}(z, z^{\prime}),
\quad \sharp=-, \pm.
\label{eqn:V}
\end{equation}
The background-background potential energies are then 
given by
\[
E^{\sharp}_{\rm bb}
=-\frac{1}{2} \frac{N^{-}}{LW} 
\int_0^L dx \int_0^W dy \,
V^{\sharp}(z),
\quad \sharp=-, \pm.
\]

The probability weights for particle configuration $\z$
at the inverse temperature $\beta >0$ are then
given by the Boltzmann factors corresponding to
the total energies,
\begin{align}
Q^{\sharp}_{\rm plasma}(\z) &=
Q^{\sharp}_{\rm plasma}(\z; N, N^{-}, \beta)
\nonumber\\
&\equiv \exp \Big[
-\beta ( E^{\sharp}_{\rm pp}(\z)
+E^{\sharp}_{\rm pb}(\z)+E^{\sharp}_{\rm bb}) \Big],
\quad \sharp=-, \pm,
\quad \z \in {\Lambda_{(L, iW)}}^N.
\label{eqn:Q_plasma}
\end{align}
These systems consisting of positively charged particles
embedded in negatively charged background
are called {\it one-component plasma models} \cite{JT96,For06,For10}.

\subsection{Calculation of potential energies}
\label{sec:calculation}

Now we evaluate $V^{\sharp}(z)$ explicitly.
By the definition (\ref{eqn:V}) with
(\ref{eqn:Phi-0}), (\ref{eqn:Phipm0}), (\ref{eqn:Phi-}),
and (\ref{eqn:Phipm}), 
\begin{align*}
V^{-}(z) &=
\frac{N_-}{LW} \Re I^{-}(z)
- 3N^{-} \log (\eta(\tau)) -N^{-} \log(2 \pi/L),
\nonumber\\
V^{\pm}(z) &=
\frac{N_-}{LW} \left( \Re I^{+}(z)+\Re I^{-}(z) - \frac{1}{2} \Re I^{0} \right)
- 3N^{-} \log (\eta(\tau)) -N^{-} \log(2 \pi/L)
\nonumber\\
& \quad
- \frac{1}{2} N^{-} \log(|\vartheta_1(2 z/L; \tau)|), 
\end{align*}
where
\begin{align*}
I^{\pm}(z) &\equiv \int_0^L dx^{\prime} \int_0^W dy^{\prime} \,
\log (\vartheta_1( (z \pm z^{\prime})/L; \tau) ),
\nonumber\\
I^{0} &\equiv \int_0^L dx^{\prime} \int_0^W dy^{\prime} \,
\log (\vartheta_1(2 z^{\prime}/L; \tau) ).
\end{align*}

As shown in Appendix \ref{sec:appendixB}, we can evaluate that
\begin{align}
I^{-}(z) &=LW \log(\eta(\tau))
+\pi \left( y-\frac{W}{2} \right)^2
+ \frac{\pi W^2}{12} 
- \pi i ( 2xy - Wx - 2Ly + LW),
\label{eqn:I-}
\\
I^{+}(z) &= LW \log(\eta(\tau))
+ \pi y^2 +\pi W y + \frac{\pi}{3} W^2 -\pi i (2xy+Wx),
\label{eqn:I+}
\\
I^0 &= LW \log(\eta(\tau)) + \frac{13 \pi}{12} W^2
- \pi i LW.
\label{eqn:I0}
\end{align}
Hence we have
\begin{align*}
V^{-}(z)
&= -2 N^{-} \log(\eta(\tau)) - N^{-} \log(2\pi/L)
+ \frac{\pi N^{-}}{LW} \left( y-\frac{W}{2} \right)^2 
+ \frac{\pi N^{-} W}{12L},
\nonumber\\
V^{\pm}(z)
&= -\frac{3}{2} N^{-} \log(\eta(\tau)) - N^{-} \log(2\pi/L)
+ \frac{2 \pi N^{-}}{LW} y^2 
+ \frac{\pi N^{-} W}{8L}
\nonumber\\
& \quad
-\frac{1}{2} N^{-}\log(|\vartheta_1(2 z/L; \tau)|), 
\end{align*}
and then we obtain
\begin{align*}
E_{\rm pb}^{-}(\z)
&=-2 N N^{-} \log(\eta(\tau)) -N N^{-} \log(2\pi/L)
+ \frac{\pi N^{-}}{LW} \sum_{j=1}^N \left( y_j-\frac{W}{2} \right)^2
+ \frac{\pi N N^{-} W}{12 L},
\nonumber\\
E_{\rm pb}^{\pm}(\z)
&=-\frac{3}{2} N N^{-} \log(\eta(\tau)) -N N^{-} \log(2\pi/L)
+ \frac{2 \pi N^{-}}{LW} \sum_{j=1}^N y_j^2
+ \frac{\pi N N^{-} W}{8 L}
\nonumber\\
& \quad
- \frac{1}{2} N^{-} \sum_{j=1}^N
\log(|\vartheta_1(2 z_j/L; \tau)|),
\end{align*}
and
\begin{align*}
E_{\rm bb}^{-}
&= (N^{-})^2 \log(\eta(\tau))+\frac{1}{2} (N^{-})^2\log(2 \pi/L)
-\frac{\pi (N^{-})^2 W}{12L},
\nonumber\\
E_{\rm bb}^{\pm}
&= (N^{-})^2 \log(\eta(\tau))
+\frac{1}{2} (N^{-})^2 \log(2 \pi/L)
-\frac{\pi (N^{-})^2 W}{8L}.
\end{align*}

The above gives the following results,
\begin{align}
E_{\rm pp}^{-}(\z)+E_{\rm pb}^{-}(\z)+E_{\rm bb}^{-}
&= - \log \left(
\prod_{1 \leq j < k \leq N} |
\vartheta_1((z_k-z_j)/L; \tau) | \right)
+ \frac{\pi N^{-}}{LW} \sum_{j=1}^N \left( y_j-\frac{W}{2} \right)^2
\nonumber\\
& \quad +\frac{1}{2} \{
3N(N-1)-4N N^{-} + 2 (N^{-})^2 \}
\log(\eta(\tau))
\nonumber\\
& \quad + \frac{1}{2} \{
N(N-1)-2 N N^{-} +(N^{-})^2 \} 
\log(2\pi/L)
\nonumber\\
& \quad + N^{-}(N-N^{-}) \frac{\pi W}{12L},
\label{eqn:E_total_-}
\end{align}
and
\begin{align}
E_{\rm pp}^{\pm}(\z)+E_{\rm pb}^{\pm}(\z)+E_{\rm bb}^{\pm}
&= - \log \left(
\prod_{1 \leq j < k \leq N} 
|\vartheta_1( (z_k-z_j)/L; \tau)| 
|\vartheta_1( (z_k+z_j)/L; \tau)| \right)
\nonumber\\
& \quad
+ \frac{1}{2} \{ (N-1)-N^{-} \} 
\log \left( \prod_{j=1}^N | \vartheta_1(2 z_j/L; \tau)| \right)
+ \frac{2 \pi N^{-}}{LW} \sum_{j=1}^N y_j^2
\nonumber\\
& \quad +\frac{1}{2} \{
3N(N-1)- 3 N N^{-} + 2 (N^{-})^2 \}
\log(\eta(\tau))
\nonumber\\
& \quad + \frac{1}{2} \{
N(N-1)-2 N N^{-} +(N^{-})^2 \} 
\log(2\pi/L)
\nonumber\\
& \quad + N^{-}(N-N^{-}) \frac{\pi W}{8L}.
\label{eqn:E_total_pm}
\end{align}

\subsection{Solvability conditions}
\label{sec:solvability}

First we review the argument given by Forrester \cite{For06}
for the first system with the two-point potential function $\Phi^{-}$.
In (\ref{eqn:E_total_-}), put
\[
N^{-}=N=\cN^{\AN}.
\]
This neutralizes the system and the last term in 
the RHS of (\ref{eqn:E_total_-}) vanishes. 
The probability weight (\ref{eqn:Q_plasma}) becomes
\begin{align*}
Q^{-}_{\rm plasma}(\z; N, \cN^{\AN}, \beta)
&=(2\pi/L)^{\beta N/2}
\eta(\tau)^{-\beta N(N-3)/2}
\nonumber\\
& \quad \times
\exp \left[ - \frac{\beta \pi N}{LW}
\sum_{j=1}^N \left( y_j- \frac{W}{2} \right)^2 \right]
\prod_{1 \leq j < k \leq N}
| \vartheta_2( (z_k-z_j)/L; \tau)|^{\beta}.
\end{align*}
Then we set
\[
\beta=2,
\]
and perform the following
transformation of the weight from
$Q^{-}_{\rm plasma}$ to
$\widehat{Q}^{-}_{\rm plasma}$,
\begin{align}
&\widehat{Q}^{-}_{\rm plasma}(\z; N, \cN^{\AN}, \beta=2)
\nonumber\\
& \quad 
=\left| \vartheta_{\widetilde{s}(N)} 
\left( \sum_{k=1}^N 
\left( \frac{z_k}{L}- \frac{L+iW}{2L} \right); \tau \right) \right|^2
Q^{-}_{\rm plasma}(\z; N, \cN^{\AN}, \beta=2),
\label{eqn:transform}
\end{align}
where $\widetilde{s}(N)=0$, if $N$ is even, and
$\widetilde{s}(N)=1$, if $N$ is odd. 

We can prove the following identities \cite{For06}.
(The proof is given in Appendix \ref{sec:appendixC}.) 

\begin{lem}
\label{thm:identities}
Let $s(N)=0, \widetilde{s}(N)=0$, if $N$ is even, and
$s(N)=3, \widetilde{s}(N)=1$, if $N$ is odd. 
Then the following equalities hold,
\begin{align}
&\exp \left[ - \frac{2 \pi N}{LW} \sum_{j=1}^N
\left( y_j-\frac{W}{2} \right)^2 \right]
\left| \vartheta_{\widetilde{s}(N)} 
\left( \sum_{k=1}^N \left( \frac{z_k}{L}
- \frac{L+iW}{2L} \right) ; \tau \right) \right|^2
\nonumber\\
& \qquad
=\exp \left( - \frac{2 \pi N}{LW} \sum_{j=1}^N y_j^2 \right)
\left| \vartheta_{s(N)} \left( \sum_{k=1}^N \frac{z_k}{L}; \tau \right) \right|^2.
\label{eqn:identities}
\end{align}
\end{lem}
\vskip 0.3cm
\noindent

Then we arrive at the following equality,
\begin{equation}
\widehat{Q}^{-}_{\rm plasma}(\z; N, \cN^{\AN}, \beta=2)
= c^{\AN}(L, \tau)Q^{\AN}(\z),
\label{eqn:QQ_A1}
\end{equation}
with
\begin{equation}
c^{\AN}(L, \tau)=(2\pi/L)^N \eta(\tau)^{-N(N-3)},
\label{eqn:QQ_A2}
\end{equation}
where $Q^{\AN}(\z)$ is given by (\ref{eqn:QRN1})
for $\RN=\AN$.
As stated in Proposition \ref{thm:totally_elliptic},
the probability weight 
$Q^{\AN}(\z)$ 
is doubly periodic with respect to
the $N$-particle configuration $\z \in \C^N$,
and hence so is \\
$\widehat{Q}^{-}_{\rm plasma}(\z; N, \cN^{\AN}, \beta=2)$ 
by the proportionality (\ref{eqn:QQ_A1}). 
Moreover, as shown in Section \ref{sec:DPP},
this probability weight can be well normalized
and the point process $\Xi^{\AN}$ is proved to be 
determinantal governed by the correlation kernel
$K^{\AN}$ given by (\ref{eqn:K1})
with $\{M^{\AN}_n(z) \}_{n=1}^N$ defined in the first line of (\ref{eqn:MA_D1}). 
Although Forrester did not give the correlation kernel
for this DPP explicitly,
he claimed the fact that if and only if we perform
the transformation (\ref{eqn:transform}) in the probability weight
at $\beta=2$, the one-component plasma model
becomes exactly solvable in the sense that
all correlation functions are explicitly given by
determinants generated by $K^{\AN}$.
The origin of this key transformation (\ref{eqn:transform})
is found in the Macdonald denominator formula
(\ref{eqn:Macdonald3}) for $\RN=\AN$
given by Rosengren and Schlosser \cite{RS06}.
(Note that Forrester \cite{For06,For10} proved the equivalent equalities
to the Macdonald denominator formula for
$\RN=\AN$ independently of \cite{RS06}. 
See Remark 8 in \cite{Kat18}.)

Now we consider the second system with the two-point
potential function $\Phi^{\pm}$.
We will report two cases which give us the 
exactly solvable plasma models.

\vskip 0.3cm
\noindent
\underline{\bf Type $\DN$}
\vskip 0.3cm

In (\ref{eqn:E_total_pm}), first we set
\[
N^{-}=N-1=\frac{\cN^{\DN}}{2}.
\]
In this setting, the second term in the RHS of (\ref{eqn:E_total_pm})
vanishes.
Since neutralization in not achieved, $N > N^{-}$,
the last term in the RHS of (\ref{eqn:E_total_pm}) remains positive.
However, if we set $\beta=2$ again, we obtain the
following equalities,
\begin{align}
& Q^{\pm}_{\rm plasma}(\z; N, \cN^{\DN}/2, \beta=2)
\nonumber\\
& \quad 
= c^{\DN}(L, \tau) \exp \left(
-\frac{2 \pi \cN^{\DN}}{LW}
\sum_{j=1}^N y_j^2 \right)
| W^{\DN}(\z/L; \tau) |^2
\nonumber\\
& \quad
= c^{\DN}(L, \tau) Q^{\DN}(\z)
\label{eqn:QQ_D1}
\end{align}
with
\begin{equation}
c^{\DN}(L, \tau)
=(2 \pi/L)^{N-1} \eta(\tau)^{-2(N-1)^2}
e^{(N-1) \tau \pi i /L}.
\label{eqn:QQ_D2}
\end{equation}

\vskip 0.3cm
\noindent
\underline{\bf Type $\CN$}
\vskip 0.3cm

Next we set
\[
N^{-}=N+1=\frac{\cN^{\CN}}{2}
\]
in (\ref{eqn:E_total_pm}).
The system is charged $-1$, and 
if we set $\beta=2$, the
following equalities are established, 
\begin{align}
& Q^{\pm}_{\rm plasma}(\z; N, \cN^{\CN}/2, \beta=2)
\nonumber\\
& \quad 
= c^{\CN}(L, \tau) \exp \left(
-\frac{2 \pi \cN^{\CN}}{LW}
\sum_{j=1}^N y_j^2 \right)
| W^{\CN}(\z/L; \tau) |^2
\nonumber\\
& \quad
= c^{\CN}(L, \tau) Q^{\CN}(\z)
\label{eqn:QQ_C1}
\end{align}
with
\begin{equation}
c^{\CN}(L, \tau)
=(2 \pi/L)^{N-1} \eta(\tau)^{-2(N^2-N+1)}
e^{-(N+1) \tau \pi i/L}.
\label{eqn:QQ_C2}
\end{equation}

The above results are summarized as follows.

\begin{thm}
\label{thm:solvable_plasma}
In the following three cases, the one-component
plasma model in $\Lambda_{(L, iW)} \subset \C$ becomes
exactly solvable in the sense that
the particle configuration $\z \in {\Lambda_{(L, iW)}}^N$
is given by DPP.

\begin{description}
\item{\rm (1)} \quad
The case such that the two-point potential is 
$\Phi^{-}$ given by (\ref{eqn:Phi-}),
$N^{-}=\cN^{\AN}$, $\beta=2$,
and the transform (\ref{eqn:transform}) is performed.
The correlation kernel is given by $K^{\AN}$.
The system is neutral.

\item{\rm (2)} \quad
The case such that the two-point potential is 
$\Phi^{\pm}$ given by (\ref{eqn:Phipm}),
$N^{-}=\cN^{\CN}/2$, and $\beta=2$.
The correlation kernel is given by $K^{\CN}$.
The system is negatively charged by unit, $-1$.

\item{\rm (3)} \quad
The case such that the two-point potential is 
$\Phi^{\pm}$ given by (\ref{eqn:Phipm}),
$N^{-}=\cN^{\DN}/2$, and $\beta=2$.
The correlation kernel is given by $K^{\DN}$.
The system is positively charged by unit, $+1$. 
\end{description}
\end{thm}
\vskip 0.3cm
\noindent{\bf Remark 8} \
As mentioned in Remark 4 in Section \ref{sec:infinite},
the Ginibre-like DPP of type $C$, 
$(\Xi^{C}, \cP^{C}_{{\rm Ginibre}, \rho})$,
realizes the Mittag--Leffler field with an 
inserted point charge of strength $c=1$ at the origin.
The inserted charge is repulsive \cite{AKS18} and hence
the density profile of other particles behaves
as (\ref{eqn:rho_C_0}) with (\ref{eqn:rho_inf_CD}). 
In the above one-component plasma model,
this inserted charge is considered to be negative
added to the negatively charged background of the system.
Similarly, the Ginibre-like DPP of type $D$, 
$(\Xi^{D}, \cP^{D}_{{\rm Ginibre}, \rho})$, 
can be interpreted as the Mittag--Leffler field
with an inserted point charge of $c=-1$ at the origin,
which is attractive to the system \cite{AKS18}.
Hence we see (\ref{eqn:rho_D_0}) with (\ref{eqn:rho_inf_CD}). 
We think that in the corresponding one-component
plasma mode there is 
a deficit of negative charge by unit in the
background.

\subsection{Relationship to Gaussian free field on a torus}
\label{sec:GFF}

We define the partition functions for the present three types of
exactly solvable one-component plasma models as
\[
Z^{\RN}_{\rm plasma}
= \frac{1}{N!} \int_{{\Lambda(L, iW)}^N}
\widetilde{Q}^{\RN}_{\rm plasma}(\z; N, N^{-}, \beta=2) d \z,
\]
with
\[
\widetilde{Q}^{\RN}_{\rm plasma}(\, \cdot \, ; \, \cdot \, , N^{-}, \, \cdot \,)
= \begin{cases}
\widehat{Q}^{-}_{\rm plasma}(\, \cdot \, ; \, \cdot \, , \cN^{\AN}, \, \cdot \,)
&\mbox{for $\RN=\AN$},
\cr
Q^{\pm}_{\rm plasma}
(\, \cdot \, ; \, \cdot \, , \cN^{\RN}/2, \, \cdot \,)
&\mbox{for $\RN=\CN, \DN$}.
\end{cases}
\]
By the equalities (\ref{eqn:QQ_A1}), (\ref{eqn:QQ_D1}), and
(\ref{eqn:QQ_C1}) 
with (\ref{eqn:QQ_A2}), (\ref{eqn:QQ_D2}) and (\ref{eqn:QQ_C2}),
we have the equalities
\[
Z^{\RN}_{\rm plasma}
= c^{\RN}(L, \tau) Z^{\RN},
\quad \RN=\AN, \CN, \DN,
\]
where $Z^{\RN}$ are given in Lemma \ref{thm:Qint}.
Hence we obtain the following exact formulas
for the solvable plasma models,
\begin{align}
Z^{\AN}_{\rm plasma}
&= \left( 2 \pi^2 \frac{LW}{N} \right)^{N/2} \eta(\tau)^2,
\nonumber\\
Z^{\CN}_{\rm plasma}
&= \left( 2^2 \pi^2 \frac{LW}{N+1} \right)^{N/2}
\frac{L}{2 \pi} e^{- (N+1) \tau \pi i/4}
\eta(\tau)^{-2},
\nonumber\\
Z^{\DN}_{\rm plasma}
&= \left( 2^2 \pi^2 \frac{LW}{N-1} \right)^{N/2}
\frac{L}{2^3 \pi} e^{(N-1) \tau \pi i/4}
\eta(\tau)^{-2}.
\label{eqn:exact_partition_functions}
\end{align}

The free energy per particle multiplied by the
inverse temperature $\beta$ is defined by
\[
F^{\RN}_{\rm plasma}(\beta) = \beta f^{\RN}_{\rm plasma}
\equiv -\frac{1}{N} \log Z^{\RN}_{\rm plasma},
\quad \RN=\AN, \CN, \DN.
\]
The exact formulas (\ref{eqn:exact_partition_functions})
give the following,
\begin{align*}
F^{\AN}_{\rm plasma}(\beta=2)
&= F^{A}_0+ \frac{1}{N} F^{A}_1,
\nonumber\\
F^{\CN}_{\rm plasma}(\beta=2)
&= F^{C}_0  - \frac{\log N}{2N} + \frac{1}{N} F^{C}_1
+{\rm O}(N^{-2}),
\nonumber\\
F^{\DN}_{\rm plasma}(\beta=2)
&= F^{D}_0  - \frac{\log N}{2N} + \frac{1}{N} F^{D}_1
+{\rm O}(N^{-2})
\end{align*}
with
\[
F^{\sharp}_0 = \begin{cases}
\displaystyle{
\frac{1}{2} \log \left( \frac{\rho}{2 \pi^2} \right)
},
& \quad \sharp=A,
\cr
& \cr
\displaystyle{
\frac{1}{2} \log \left( \frac{\rho}{4 \pi^2} \right)
-\frac{\pi}{4} \Im \tau
},
& \quad \sharp=C,
\cr
& \cr
\displaystyle{
\frac{1}{2} \log \left( \frac{\rho}{4 \pi^2} \right)
+\frac{\pi}{4} \Im \tau
},
& \quad \sharp=D,
\end{cases}
\]
and
\begin{equation}
F^{\sharp}_1=\begin{cases}
\displaystyle{
-\log(\eta(\tau)^2)
},
& \sharp=A,
\cr
& \cr
\displaystyle{
\log\{2 \sqrt{\pi \Im \tau} \eta(\tau)^2 \}
+\frac{1}{2} \log(\pi \rho)
-\left( \frac{\pi}{4} \Im \tau -\frac{1}{2} \right)
},
& \sharp=C,
\cr
& \cr
\displaystyle{
\log\{2 \sqrt{\pi \Im \tau} \eta(\tau)^2 \}
+\frac{1}{2} \log(2^4 \pi \rho)
-\left(\frac{\pi}{4} \Im \tau +\frac{1}{2}\right)
},
& \sharp=D,
\cr
\end{cases}
\label{eqn:F_correction}
\end{equation}
where $\rho=N/(LW)$.

As pointed out by Forrester \cite{For06},
it was shown by Cardy \cite{Car90} that, 
if we ignore the zero mode, the partition function of
the Gaussian free field (GFF) on a torus 
with the modular parameter $\tau \in \H$ is given by
\[
Z^{(k \not= 0)}_{\rm GFF}(\tau)
=\frac{1}{\eta(\tau) \eta(-\overline{\tau})}=\frac{1}{|\eta(\tau)|^2}, 
\]
while if we treat the zero mode with appropriate regularization,
it is determined in the form with a factor
proportional to the inverse square root of $\Im \tau$,
\[
Z_{\rm GFF}(\tau)=
\frac{1}{2 \sqrt{\pi \Im \tau} |\eta(\tau)|^2},
\]
(see the end of Section 3 of \cite{KM17}).
In the present setting, $\tau$ is pure imaginary in $\H$
and hence $\eta(\tau) \in \R$.
Forrester clarified the following equality 
with a negative sign \cite{JT96,For06}, 
\[
F^{A}_1= - F_{\rm GFF}^{(k \not=0)}(\tau)
\quad \mbox{with 
$ F_{\rm GFF}^{(k \not=0)}(\tau) \equiv 
-\log Z^{(k \not=0)}_{\rm GFF}(\tau)= \log( \eta(\tau)^2)$}.
\]
On the other hand, here we state that 
$F^{C}_1$ and $F^{D}_1$ include the term
\[
F_{\rm GFF}(\tau)
\equiv -\log Z_{\rm GFF}(\tau)
=\log\{ 2 \sqrt{\pi \Im \tau} \eta(\tau)^2\}
\]
without change of sign.
Dedekind's $\eta$ function satisfies the functional equation, 
$\eta (-1/\tau)
=(-i \tau)^{1/2} \eta(\tau)$ \cite{NIST10}.
This implies the equality
\[
\sqrt{\Im \tau} |\eta(\tau)|^2
=\sqrt{\Im(-1/\tau)} |\eta(-1/\tau)|^2,
\]
that is, the term $\log \{2 \sqrt{ \pi \Im \tau} |\eta(\tau)|^2\}$ is
invariant under the transformation $\tau \to -1/\tau$.
Since we consider the case that
$\tau=iW/L$, this is the invariance under the
change of the aspect ratio 
$W/L \to L/W$. 

\SSC
{Concluding Remarks} \label{sec:remarks}

In the present paper, we proposed a new parameterization (\ref{eqn:setting1})
of the seven families of
$\RN$-theta functions given by Rosengren and Schlosser \cite{RS06}
associated with the seven 
types of irreducible reduced affine root systems,
$W^{\RN}(\bxi)$, $\bxi=(\xi_1, \dots, \xi_N) \in \C^N$,
$\RN=\AN$, $\BN, \BNv, \CN, \CNv, \BCN, \DN$,
$N \in \N$. 
Then we proved the orthogonality relations 
with respect to the double integrals over the
fundamental domain $\Lambda_{(L, iW)} \subset \C$
of these $\RN$-theta functions (Theorem \ref{thm:orthogonality_z}).
The orthogonality relations are essential to obtain
new families of elliptic DPPs,
$(\Xi^{\RN}, \bP^{\RN})$, on a complex plane $\C$. 
The connection to one-component plasma models 
was able to be discussed based on these orthogonality relations,
but the results are limited to the three types
$\AN$, $\CN$, and $\DN$.
Since the partition functions of our elliptic DPPs, 
$(\Xi^{\RN}, \bP^{\RN})$, are explicitly evaluated
for all seven types in Lemma \ref{thm:Qint}, 
other plasma models
are desired whose particle sections
realize $(\Xi^{\RN}, \bP^{\RN})$ for
$\RN \not=\AN, \CN, \DN$.
Generalization of the present results to
two-component plasma models \cite{For90,JT96,For06}
(see also Section 2.2 in \cite{Sos00})
will be a challenging future problem.

In Theorem \ref{thm:scaling_limits}
the DPPs with an infinite number of points are obtained.
They are defined on $\C$ having periodicity with period $i W$.
Appearance of such strip structures, 
$\cD_{(n+1) W} \equiv \{z \in \C : n W \leq \Im z < (n+1) W\}$, 
$n \in \Z$, on $\C$ is due to the scaling limit
$N \to \infty, L \to \infty$ with
constant density $\rho=N/(LW)$ and constant $W$.
Statistics of point processes in a strip $\cD_W \subset \C$ is
an interesting and important topic in mathematics
\cite{PW34,Fel13,AHKR18}. 
In the further limit $W \to \infty$ with constant $\rho$,
three types of infinite-dimensional  DPPs are obtained. 
One of them is identified with the Ginibre point process, 
which is uniform on $\C$ and realized as the eigenvalue distribution
of complex Gaussian random matrices \cite{Gin65,HKPV09,Shi15}.
Other two point processes are rotationally symmetric around the origin, 
but non-uniform in $\C$,
which will provide new examples of the Mittag--Leffler fields
studied in \cite{AK13,AKM18,AKS18}, see Remark 5 in 
Section \ref{sec:infinite}.
As mentioned in Remarks 2 and 6 also in Section \ref{sec:infinite},
the four types of correlation kernels
$\cK^{R}_{W, \rho}, R=A, B, C, D$ given 
in Proposition \ref{thm:infinite_systems1}
and the three types of Ginibre-like kernels,
$\cK^{R}_{{\rm Ginibre}, \rho}, R=A, C, D$
can be regarded as {\it reproducing and projection kernels} 
associated with properly defined unitary transformations
in appropriate Hilbert spaces.
Systematic study is now on progress to characterize
correlation kernels for finite and infinite DPPs
not only in $\R$ or $\C$ but also on the 
higher dimensional manifolds \cite{KS}.
See also \cite{HSSS17,MOC18} and references therein.
Random matrix ensembles which give these generalized DPPs
as eigenvalue distributions should be also studied
(see Remark 3 in Section \ref{sec:infinite}). 

We reported the relationship between 
the elliptic DPPs
and the GFF on a torus following the argument given by Forrester \cite{For06}.
It is not yet known whether 
the relationship found in the large-$N$ expansions
of the free energies means more direct connections
between some limit systems of 
the present elliptic DPPs and random fields related to GFF \cite{She07}
in the level of probability laws and geometrical structures.
In the systems of types C and D, the difference from type A
of signs for the $\eta(\tau)$-terms
found in  (\ref{eqn:F_correction})
suggests the boundary condition of the corresponding random field
may be different from the Dirichlet boundary condition. 
The regularized elliptic determinantal systems should be 
invariant under the transformation $\tau \to -1/\tau$,
and hence they will be invariant under the modular group SL(2, $\Z$)
of transformations of the form,  
$\tau \to (a \tau + b)/(c \tau +d)$,
$a, b, c, d \in \Z, ad-bc = 1$.

\vskip 0.5cm
\noindent{\bf Acknowledgements} \,
On sabbatical leave from Chuo University, 
this study was done in 
Fakult\"{a}t f\"{u}r Mathematik, Universit\"{a}t Wien,
in which the present author thanks Christian Krattenthaler very much
for his hospitality. 
The author thanks 
Peter John Forrester for useful comments on 
the two-dimensional Coulomb gas models.
He also expresses his gratitude to Michael Schlosser 
and Tomoyuki Shirai for valuable discussion 
concerning the present study.
This work was supported by
the Grant-in-Aid for Scientific Research (C) (No.26400405),
(B) (No.18H01124), and 
(S) (No.16H06338) 
of Japan Society for the Promotion of Science.
It was also supported by 
the Research Institute for Mathematical Sciences (RIMS),
a Joint Usage/Research Center located in Kyoto University.
The author thanks Naotaka Kajino, Takashi Kumagai, 
and Daisuke Shiraishi for organizing the very fruitful workshop, 
`RIMS Research Project: Gaussian Free Fields and Related Topics',
held in 18-21 September 2018 at RIMS.


\appendix
\SSC{The Jacobi Theta Functions}
\label{sec:appendixA}
Let
\[
z=e^{v \pi i}, \quad q=e^{\tau \pi i},
\]
where $v, \tau \in \C$ and $\Im \tau > 0$. 
The Jacobi theta functions are defined as follows \cite{WW27,NIST10}, 
\begin{align}
\vartheta_0(v; \tau) &= 
-i e^{(v+\tau/4) \pi i} \vartheta_1 \left( v + \frac{\tau}{2}; \tau \right)=
\sum_{n \in \Z} (-1)^n q^{n^2} z^{2n} 
=1+ 2 \sum_{n=1}^{\infty}(-1)^n e^{\tau \pi i n^2} \cos(2 n \pi v),
\nonumber\\
\vartheta_1(v; \tau) &= i \sum_{n \in \Z} (-1)^n q^{(n-1/2)^2} z^{2n-1}
=2 \sum_{n=1}^{\infty} (-1)^{n-1} e^{\tau \pi i (n-1/2)^2} \sin\{(2n-1) \pi v\},
\nonumber\\
\vartheta_2(v; \tau) 
&= \vartheta_1 \left( v+ \frac{1}{2}; \tau \right)=
\sum_{n \in \Z} q^{(n-1/2)^2} z^{2n-1}
=2 \sum_{n=1}^{\infty} e^{\tau \pi i (n-1/2)^2} \cos \{(2n-1) \pi v\},
\nonumber\\
\vartheta_3(v; \tau) 
&= 
e^{(v+\tau/4) \pi i } \vartheta_1 \left( v+\frac{1+\tau}{2}; \tau \right)=
\sum_{n \in \Z} q^{n^2} z^{2n}
=1 + 2 \sum_{n=1}^{\infty} e^{\tau \pi i n^2} \cos (2 n \pi v).
\label{eqn:theta}
\end{align}
(Note that the present functions 
$\vartheta_{\mu}(v; \tau), \mu=1,2,3$ are denoted by
$\vartheta_{\mu}(\pi v,q)$,
and $\vartheta_0(v;\tau)$ by $\vartheta_4(\pi v,q)$ in \cite{WW27}.)
For $\Im \tau >0$, $\vartheta_{\mu}(v; \tau)$, $\mu=0,1,2,3$
are holomorphic for $|v| < \infty$.
The parity with respect to $v$ is given by
\begin{equation}
\vartheta_1(-v; \tau)=-\vartheta_1(v; \tau),
\quad
\vartheta_{\mu}(-v; \tau)=\vartheta_{\mu}(v; \tau), \quad
\mu=0,2,3,
\label{eqn:even_odd}
\end{equation}
and they have the quasi-double-periodicity; 
\begin{align}
\vartheta_{\mu}(v+1; \tau) &=
\begin{cases}
\vartheta_{\mu}(v; \tau), & \mu=0, 3,
\cr
- \vartheta_{\mu}(v; \tau), & \mu=1, 2,
\end{cases}
\label{eqn:quasi_periodic1}
\\
\vartheta_{\mu}(v+\tau; \tau) &=
\begin{cases}
-e^{-(2v+\tau) \pi i } \vartheta_{\mu}(v; \tau),
& \mu=0, 1,
\cr
e^{-(2v+\tau) \pi i } \vartheta_{\mu}(v; \tau),
& \mu=2,3.
\end{cases}
\label{eqn:quasi_periodic2}
\end{align}
By the definition (\ref{eqn:theta}), 
when $\tau \in \H$. 
\begin{align*}
& \vartheta_1(0; \tau)=\vartheta_1(1; \tau)=0, \qquad
\vartheta_1(x; \tau) > 0, \quad x \in (0,1),
\nonumber\\
& \vartheta_2(-1/2; \tau)=\vartheta_2(1/2; \tau)=0, \qquad
\vartheta_2(x; \tau) > 0, \quad x \in (-1/2, 1/2),
\nonumber\\
& \vartheta_0(x; \tau) > 0, \quad \vartheta_3(x; \tau) > 0, \quad x \in \R.
\end{align*}

We see the asymptotics
\begin{align}
& \vartheta_0(v; \tau) \simeq 1, \quad
\vartheta_1(v; \tau) \simeq 2 e^{\tau \pi i/4} \sin (\pi v), \quad
\vartheta_2(v; \tau) \simeq 2 e^{\tau \pi i/4} \cos(\pi v), \quad
\vartheta_3(v; \tau) \simeq 1,
\nonumber\\
&\qquad \qquad \qquad \mbox{in} \quad
\Im \tau \to + \infty \quad
({\it i.e.}, \quad q=e^{\tau \pi i} \to 0).
\label{eqn:theta_asym}
\end{align}

The Jacobi theta function $\vartheta_1$ defined by
(\ref{eqn:theta}) has the following infinite-product expressions, 
\begin{eqnarray}
\vartheta_1(v ; \tau)
&=& -i q^{1/4} z \prod_{j=1}^{\infty} 
\Big\{ (1-q^{2j} z^2) (1-q^{2j-2}/z^2) (1-q^{2j}) \Big\}
\nonumber\\
&=& 2 q^{1/4} \sin(\pi v)
\prod_{j=1}^{\infty}
\Big\{ (1-2 q^{2j} \cos(2 \pi v) + q^{4j}) (1-q^{2j}) \Big\}. 
\label{eqn:theta_product_1}
\end{eqnarray}
The following is also known,
\begin{equation}
\left. \frac{\partial \vartheta_1(v; \tau)}{\partial v} \right|_{v=0}
=2 \pi \eta(\tau)^3,
\label{eqn:diff_theta1}
\end{equation}
where $\eta(\tau)$ is the Dedekind modular function (\ref{eqn:Dedekind1}).

The following functional equalities are known as
Jacobi's imaginary transformations \cite{WW27,NIST10},
\begin{align}
\vartheta_0(v; \tau)
&= e^{\pi i/4} \tau^{-1/2} e^{-\pi i v^2/\tau}
\vartheta_2 \left( \frac{v}{\tau}; - \frac{1}{\tau} \right),
\nonumber\\
\vartheta_1(v; \tau)
&= e^{3 \pi i/4} \tau^{-1/2} e^{-\pi i v^2/\tau}
\vartheta_1 \left( \frac{v}{\tau}; - \frac{1}{\tau} \right),
\nonumber\\
\vartheta_2(v; \tau)
&= e^{\pi i/4} \tau^{-1/2} e^{-\pi i v^2/\tau}
\vartheta_0 \left( \frac{v}{\tau}; - \frac{1}{\tau} \right), 
\nonumber\\
\vartheta_3(v; \tau)
&= e^{\pi i/4} \tau^{-1/2} e^{-\pi i v^2/\tau}
\vartheta_3 \left( \frac{v}{\tau}; - \frac{1}{\tau} \right). 
\label{eqn:Jacobi_imaginary}
\end{align}

\SSC{Evaluation of Integrals}
\label{sec:appendixB}
\subsection{Integral $I^-(z)$}
\label{sec:I-}

Consider the integral
\begin{align*}
I^{-}(z) &=\int_0^L dx^{\prime} \int_0^W dy^{\prime} \,
\log (\vartheta_1((z-z^{\prime})/L; \tau))
\nonumber\\
&= \int_0^L dx^{\prime} \int_0^W dy^{\prime} \,
\log \left\{ \vartheta_1 \left( \frac{x-x^{\prime}}{L}
+i \frac{y-y^{\prime}}{L}; \tau \right) \right\}.
\end{align*}
By the product formula (\ref{eqn:theta_product_1}),
\[
I^{-}(z)=\sum_{j=1}^5 I^{-}_j
\]
with
\begin{align*}
I^{-}_1 &= \int_0^L dx^{\prime} \int_0^W dy^{\prime} \,
\left(- \frac{\pi \Im \tau}{4} \right)
=-\frac{\pi}{4}  \Im \tau LW = - \frac{\pi}{4} W^2,
\nonumber\\
I^{-}_2 &=\int_0^L dx^{\prime} \int_0^W dy^{\prime} \,
\log \left[ 2 \sin \left[
\frac{\pi}{L} \left\{ (x-x^{\prime})+ i (y-y^{\prime}) \right\} \right] \right],
\nonumber\\
I^{-}_3 &= \sum_{n=1}^{\infty}
\int_0^L dx^{\prime} \int_0^W dy^{\prime} \,
\log(1-q^{2n} e^{2 \pi i (x-x^{\prime})/L -  2\pi (y-y^{\prime})/L}),
\nonumber\\
I^{-}_4 &= \sum_{n=1}^{\infty}
\int_0^L dx^{\prime} \int_0^W dy^{\prime} \,
\log(1-q^{2n} e^{-2 \pi i (x-x^{\prime})/L +  2\pi (y-y^{\prime})/L}),
\nonumber\\
I^{-}_5 &= \int_0^L dx^{\prime} \int_0^W dy^{\prime} \, 
\sum_{n=1}^{\infty} \log(1-q^{2n})
= LW \sum_{n=1}^{\infty} \log(1-q^{2n}),
\end{align*}
where $q=e^{\tau \pi i}$.
By the definition of Dedekind's function
(\ref{eqn:Dedekind1}), we readily see that
\[
I^{-}_5 = LW 
\left(\frac{\pi}{12} \Im \tau
+ \log (\eta(\tau)) \right)
= \frac{\pi}{12} W^2
+LW \log (\eta(\tau)).
\]

Note that 
\[
q^{2n}e^{\pm 2 \pi i (x-x^{\prime})/L \mp 2 \pi (y-y^{\prime})/L}
= \exp \left[- \frac{2 \pi}{L}
\{ nW \pm (y-y^{\prime}) \} \right]
e^{\pm 2 \pi i (x-x^{\prime})/L}.
\]
Since $0 \leq y, y^{\prime} \leq W$, $|y-y^{\prime}| \leq W$
and thus for $n \geq 1$,
$n W \pm (y-y^{\prime}) \geq 0$.
Then we have the expansion for $I^{-}_3$ as
\[
I^{-}_3
= - \sum_{n=1}^{\infty} \sum_{k=1}^{\infty}
\frac{q^{2nk}}{k}
\int_0^L e^{2 \pi i (x-x^{\prime})k/L} dx^{\prime}
\int_0^W e^{-2 \pi (y-y^{\prime})k/L} dy^{\prime}.
\]
For $k \geq 1$, $\int_0^L e^{2 \pi i (x-x^{\prime})k/L} dx^{\prime} =0$,
and hence $I^{-}_3=0$.
Similarly, $I^{-}_4=0$.

Now we consider the integrand of $I^{-}_2$,
\[
\log \left[ 2 \sin \left[
\frac{\pi}{L} \{ (x-x^{\prime})+i (y-y^{\prime}) \} \right] \right]
= - \frac{\pi i}{2}
+\log \Big(
e^{\pi i (x-x^{\prime})/L} e^{-\pi (y-y^{\prime})/L}
- e^{- \pi i(x-x^{\prime})/L} e^{\pi (y-y^{\prime})/L} \Big).
\]
When $y \geq y^{\prime}$, this is written as
\begin{align*}
& - \frac{\pi i}{2}
+ \log \left[
(-e^{- \pi i (x-x^{\prime})/L} e^{\pi(y-y^{\prime})/L})
(1-e^{2 \pi i (x-x^{\prime})/L} e^{-2 \pi (y-y^{\prime})/L} \right]
\nonumber\\
& \quad
= \frac{\pi i}{2}
- \frac{\pi i}{L}(x+i y) + \frac{\pi i}{L} (x^{\prime}+i y^{\prime})
-\sum_{k=1}^{\infty} \frac{1}{k} e^{2 \pi i (x-x^{\prime})k/L}
e^{-2 \pi (y-y^{\prime})k/L},
\end{align*}
and, when $y < y^{\prime}$, the above integrand is 
written as
\begin{align*}
& - \frac{\pi i}{2}
+ \log \left[
e^{\pi i (x-x^{\prime})/L} e^{-\pi(y-y^{\prime})/L}
(1-e^{-2 \pi i (x-x^{\prime})/L} e^{2 \pi (y-y^{\prime})/L} \right]
\nonumber\\
& \quad
= -\frac{\pi i}{2}
+ \frac{\pi i}{L}(x+i y) - \frac{\pi i}{L} (x^{\prime}+i y^{\prime})
-\sum_{k=1}^{\infty} \frac{1}{k} e^{-2 \pi i (x-x^{\prime})k/L}
e^{2 \pi (y-y^{\prime})k/L}.
\end{align*}
Since 
$\int_0^L dx^{\prime} e^{\pm 2 \pi i (x-x^{\prime})/L}=0$, 
$k \geq 1$, for given $(x, y) \in [0, L] \times [0, W]$, we have
\[
I^{-}_2= I^{-, <}_2+I^{-, >}_2
\]
with
\begin{align*}
 I^{-, <}_2 & \equiv
\int_0^L dx^{\prime} \int_0^y dy^{\prime} \,
\left\{ 
\frac{\pi i}{2}
- \frac{\pi i}{L}(x+i y) + \frac{\pi i}{L} (x^{\prime}+i y^{\prime}) 
\right\}
\nonumber\\
&=
\frac{\pi i}{2} Ly - \frac{\pi i}{L}(x+iy) Ly + \frac{\pi i}{L}
\left( \frac{L^2}{2} y + i L \frac{y^2}{2} \right),
\nonumber\\
 I^{-, >}_2 & \equiv
\int_0^L dx^{\prime} \int_y^W dy^{\prime} \,
\left\{ 
 -\frac{\pi i}{2}
+ \frac{\pi i}{L}(x+i y) - \frac{\pi i}{L} (x^{\prime}+i y^{\prime})
\right\}
\nonumber\\
&=
-\frac{\pi i}{2} L(W-x) + \frac{\pi i}{L}(x+iy) L (W-y)
- \frac{\pi i}{L} \left\{
\frac{L^2}{2} (W-y) + i L \left( \frac{W^2}{2}-\frac{y^2}{2} \right) \right\},
\end{align*}
and hence
\[
I^{-}_2= \pi \left( y-\frac{W}{2} \right)^2 + \frac{\pi}{4} W^2
- \pi i (2xy-Wx-2Ly+LW).
\]

Combining the above results, we obtain (\ref{eqn:I-}). 

\subsection{Integral $I^{+}(z)$}
\label{sec:I+}

Consider the integral
\begin{align*}
I^{+}(z) &=\int_0^L dx^{\prime} \int_0^W dy^{\prime} \,
\log (\vartheta_1((z+z^{\prime})/L; \tau))
\nonumber\\
&= \int_0^L dx^{\prime} \int_0^W dy^{\prime} \,
\log \left\{ \vartheta_1 \left( \frac{x+x^{\prime}}{L}
+i \frac{y+y^{\prime}}{L}; \tau \right) \right\}.
\end{align*}
By the quasi-double-periodicity (\ref{eqn:quasi_periodic2}), the integrand is equal to
\[
\pi i - \frac{2\pi i}{L}(x+x^{\prime})
+\frac{2\pi}{L}(y+y^{\prime}) - \frac{\pi W}{L}
+ \log \vartheta_1 \left(
\frac{x+x^{\prime}}{L}+i \frac{y+y^{\prime}-W}{L} ; \tau \right).
\]
By the product formula (\ref{eqn:theta_product_1}),
\[
I^{+}(z)=\sum_{j=0}^5 I^{+}_j
\]
with
\begin{align*}
I^{+}_0 &= \int_0^L dx^{\prime} \int_0^W dy^{\prime} \,
\left\{ 
\pi i - \frac{2\pi i}{L}(x+x^{\prime})
+\frac{2\pi}{L}(y+y^{\prime}) - \frac{\pi W}{L}
\right\}
\nonumber\\
&= 2 \pi y W -2 \pi i W x,
\nonumber\\
I^{+}_1 &= \int_0^L dx^{\prime} \int_0^W dy^{\prime} \,
\left(- \frac{\pi}{4} \Im \tau \right)
=-\frac{\pi}{4} \Im \tau LW = - \frac{\pi}{4} W^2,
\nonumber\\
I^{+}_2 &=\int_0^L dx^{\prime} \int_0^W dy^{\prime} \,
\log \left[ 2 \sin \left[
\frac{\pi}{L} \left\{ (x+x^{\prime})+ i (y+y^{\prime}-W) \right\} \right] \right],
\nonumber\\
I^{+}_3 &= \sum_{n=1}^{\infty}
\int_0^L dx^{\prime} \int_0^W dy^{\prime} \,
\log(1-q^{2n} e^{2 \pi i (x+x^{\prime})/L -  2\pi (y+y^{\prime}-W)/L}),
\nonumber\\
I^{+}_4 &= \sum_{n=1}^{\infty}
\int_0^L dx^{\prime} \int_0^W dy^{\prime} \,
\log(1-q^{2n} e^{-2 \pi i (x+x^{\prime})/L +  2\pi (y+y^{\prime}-W)/L}),
\nonumber\\
I^{+}_5 &= \int_0^L dx^{\prime} \int_0^W dy^{\prime} \, 
\sum_{n=1}^{\infty} \log(1-q^{2n})
= \frac{\pi}{12} W^2
+LW \log (\eta(\tau)).
\end{align*}

Note that 
\[
q^{2n}e^{\pm 2 \pi i (x+x^{\prime})/L \mp 2 \pi (y+y^{\prime}-W)/L}
= \exp \left[- \frac{2 \pi}{L}
\{ nW \pm (y+y^{\prime}-W) \} \right]
e^{\pm 2 \pi i (x+x^{\prime})/L}.
\]
Since $0 \leq y, y^{\prime} \leq W$, $|y+y^{\prime}-W| \leq W$
and thus for $n \geq 1$,
$n W \pm (y+y^{\prime}-W) \geq 0$.
Then we have the expansion for $I^{+}_3$ as
\[
I^{+}_3
= - \sum_{n=1}^{\infty} \sum_{k=1}^{\infty}
\frac{q^{2nk}}{k}
\int_0^L e^{2 \pi i (x+x^{\prime})k/L} dx^{\prime} 
\int_0^W e^{-2 \pi (y+y^{\prime}-W)k/L} dy^{\prime}.
\]
For $k \geq 1$, $\int_0^L e^{2 \pi i (x+x^{\prime})k/L} dx^{\prime} =0$,
and hence $I^{+}_3=0$.
Similarly, $I^{+}_4=0$.

Now we consider the integrand of $I^{+}_2$,
\begin{align*}
&\log \left[ 2 \sin \left[
\frac{\pi}{L} \{ (x+x^{\prime})+i (y+y^{\prime}-W) \} \right] \right]
\nonumber\\
& \qquad
= - \frac{\pi i}{2}
+\log \Big(
e^{\pi i (x+x^{\prime})/L} e^{-\pi (y+y^{\prime}-W)/L}
- e^{- \pi i(x+x^{\prime})/L} e^{\pi (y+y^{\prime}-W)/L} \Big).
\end{align*}
When $y+y^{\prime}-W \geq 0$, this can be expanded as
\[
\frac{\pi i}{2}
- \frac{\pi i}{L}(x+i y) - \frac{\pi i}{L} (x^{\prime}+i y^{\prime}) 
- \frac{\pi W}{L}
-\sum_{k=1}^{\infty} \frac{1}{k} e^{2 \pi i (x+x^{\prime})k/L}
e^{-2 \pi (y+y^{\prime}-W)k/L},
\]
and, when $y +y^{\prime}-W < 0$, 
this can be expanded as
\[
-\frac{\pi i}{2}
+ \frac{\pi i}{L}(x+i y) + \frac{\pi i}{L} (x^{\prime}+i y^{\prime})
+ \frac{\pi W}{L}
-\sum_{k=1}^{\infty} \frac{1}{k} e^{-2 \pi i (x+x^{\prime})k/L}
e^{2 \pi (y+y^{\prime}-W)k/L}.
\]
Since 
$\int_0^L e^{\pm 2 \pi i (x+x^{\prime})/L} dx^{\prime}=0$, $k \geq 1$, 
for given $(x, y) \in [0, L] \times [0, W]$, we have
\[
I^{+}_2= I^{+, >}_2+I^{+, <}_2
\]
with
\begin{align*}
I^{+, >}_2 & \equiv
\int_0^L dx^{\prime} \int_{-y+W}^W dy^{\prime} \,
\left\{ 
\frac{\pi i}{2}
- \frac{\pi i}{L}(x+i y) - \frac{\pi i}{L} (x^{\prime}+i y^{\prime}) 
- \frac{\pi W}{L}
\right\}
\nonumber\\
&=
\frac{\pi}{2} y^2 - \pi i x y,
\nonumber\\
I^{+, <}_2 & \equiv
\int_0^L dx^{\prime} \int_0^{-y+W} dy^{\prime} \,
\left\{ 
 -\frac{\pi i}{2}
+ \frac{\pi i}{L}(x+i y) + \frac{\pi i}{L} (x^{\prime}+i y^{\prime})
+\frac{\pi W}{L}
\right\}
\nonumber\\
&=
\frac{\pi}{2} y^2- \pi W y + \frac{\pi}{2} W^2 
- \pi i x (y-W),
\end{align*}
and hence 
\[
I^{+}_2= \pi y^2-\pi W y + \frac{\pi}{2} W^2
-\pi i (2 xy -Wx).
\]

Combining the above results, we obtain (\ref{eqn:I+}). 

\subsection{Integral $I^{0}$}
\label{sec:I0}

Consider the integral
\begin{align*}
I^{0} &=\int_0^L dx^{\prime} \int_0^W dy^{\prime} \,
\log (\vartheta_1(2 z^{\prime}/L; \tau))
\nonumber\\
&= \int_0^L dx^{\prime} \int_0^W dy^{\prime} \,
\log \left\{ \vartheta_1 \left( \frac{2x^{\prime}}{L}
+i \frac{2y^{\prime}}{L}; \tau \right) \right\}.
\end{align*}
By the quasi-double-periodicity (\ref{eqn:quasi_periodic2}), the integrand is equal to
\[
\pi i - \frac{4\pi i}{L}(x^{\prime}+y i ^{\prime}) - \frac{\pi W}{L}
+ \log \vartheta_1 \left(
\frac{2x^{\prime}}{L}+i \frac{2y^{\prime}-W}{L} ; \tau \right).
\]
By the similar argument to those given in the previous
two subsection, we can show that 
\[
I^{0}(z)=(\pi W^2-\pi i L W)
-\frac{\pi}{4} W^2 + I^0_2 + 
\left( \frac{\pi}{12} W^2 + LW \log(\eta(\tau)) \right),
\]
with
\begin{align*}
I^{0}_2 &= \int_0^L dx^{\prime} \int_0^W dy^{\prime} \,
\log \left[ 2 \sin \left[
\frac{\pi}{L} \left\{ (2 x^{\prime}+ i (2y^{\prime}-W) \right\} \right] \right],
\nonumber\\
&= \frac{\pi^2}{2} W^2.
\end{align*}
Then we obtain (\ref{eqn:I0}). 

\SSC{Proof of Lemma \ref{thm:identities}}
\label{sec:appendixC}

First we assume $N$ is even and put $N=2n, n \in \N$.
In this case
\begin{equation}
\vartheta_0 \left( \sum_{k=1}^N \left( 
\frac{z_k}{L}-\frac{L+iW}{2L} \right) ; \tau \right)
= \vartheta_0 \left( -\sum_{k=1}^N \frac{z_k}{L} + (1+\tau) n; \tau \right), 
\label{eqn:EqA1}
\end{equation}
where we used (\ref{eqn:even_odd}).
By the quasi-double-periodicity (\ref{eqn:quasi_periodic1}) and
(\ref{eqn:quasi_periodic2}) of $\vartheta_0(v; \tau)$,
we have the equality
\[
\vartheta_0(v+(1+\tau); \tau)=- e^{-(2v+\tau) \pi i} \vartheta_0(v; \tau).
\]
Using this equality $n$ times, RHS of (\ref{eqn:EqA1}) is written as
\[
(-1)^n \exp \left[ \pi i \left\{
\frac{N}{L} \sum_{j=1}^N x_j - (n-1) n \right\} \right]
\exp \left[ - \frac{N \pi}{L} \sum_{j=1}^N y_j - \frac{\tau \pi i N^2}{4} \right]
\vartheta_0 \left( \sum_{k=1}^N \frac{z_k}{L}; \tau \right).
\]
Since
\begin{equation}
-\frac{N \pi}{L} \sum_{j=1}^N y_j - \frac{\tau \pi i N^2}{4}
= -\frac{\pi N}{L W}
\left\{ \sum_{j=1}^N y_j^2- \sum_{j=1}^N \left( y_j-\frac{W}{2} \right)^2 \right\},
\label{eqn:equalityA}
\end{equation}
(\ref{eqn:identities}) is obtained for even $N$.

Next we assume $N$ is odd and put $N=2n+1, n \in \N$.
In this case
\begin{equation}
\vartheta_1 \left( \sum_{k=1}^N 
\left( \frac{z_k}{L}-\frac{L+iW}{2L} \right); \tau \right)
= \vartheta_1 \left( - \sum_{k=1}^N \frac{z_k}{L} + (1+\tau) n
+ \frac{1}{2}(1+ \tau); \tau \right).
\label{eqn:EqA2}
\end{equation}
By the definition (\ref{eqn:theta}),
\[
\vartheta_1 (v+(1+\tau)/2; \tau)
=e^{-(v+\tau/4) \pi i} \vartheta_3(v; \tau),
\]
and by (\ref{eqn:quasi_periodic1}) and (\ref{eqn:quasi_periodic2}),
\[
\vartheta_3(\tau+(1+\tau);\tau)
=e^{-(2v+\tau) \pi i} \vartheta_3(v; \tau).
\]
Thus RHS of (\ref{eqn:EqA2}) is written as
\[
\exp \left[ \pi i \left\{
\frac{N}{L} \sum_{j=1}^N x_j-n^2 \right\} \right]
\exp \left[ - \frac{N \pi}{L} \sum_{j=1}^N y_j
- \frac{\tau \pi i N^2}{4} \right]
\vartheta_3 \left(
\sum_{k=1}^N \frac{z_k}{L}; \tau \right).
\]
Then, through (\ref{eqn:equalityA}),
(\ref{eqn:identities}) is obtained for odd $N$.
The proof is complete. \qed


\end{document}